\documentclass[useAMS,usenatbib]{mn2e}
\usepackage{graphicx} 
\usepackage{rotating}
\title [ X-ray occultations in AGNs] {
Search for X-ray occultations in Active Galactic Nuclei}
\author [Torricelli, Pietrini, Risaliti, Salvati]
{G. Torricelli-Ciamponi$^1$\thanks{E-mail: torricel@arcetri.astro.it},  P. Pietrini$^2$, G. Risaliti$^{1,3}$, M.  Salvati$^1$
\\
$^1$ INAF - Osservatorio Astrofisico di Arcetri, L.go E. Fermi 5,
Firenze, Italy\\
$^2$Universit\`a di Firenze, Largo E.~Fermi 2, Firenze, Italy\\
$^3$ Harvard-Smithsonian Center for Astrophysics, 60 Garden St. 
Cambridge, MA 02138 USA {E-mail: grisaliti@cfa.harvard.edu}\\
}
\begin{document}

\date{Released Xxxx Xxxxx XX}

\pagerange{\pageref{firstpage}--\pageref{lastpage}} \pubyear{2002}

\maketitle

\label{firstpage}

\begin{abstract}
Recent time-resolved spectral studies of a few Active Galactic Nuclei in hard X-rays revealed occultations of the X-ray primary source probably by Broad Line Region (BLR) clouds. An important open question on the structure of the circumnuclear medium of AGN is whether this phenomenon is common, i.e. whether a significant fraction of the X-ray absorption in AGN is due to BLR clouds. Here we present the first attempt to perform this kind of analysis in a homogeneous way, on a statistically representative sample of AGN, consisting of the $\sim$40 brightest sources with long {\em XMM-Newton} and/or {\em Suzaku} observations. We describe our method, based on a simple analysis of hardness-ratio light curves, and its validation through a complete spectroscopic analysis of a few cases. We find that X-ray eclipses, most probably due to clouds at the distance of the BLR, are common in sources where the expected occultation time is compatible with the observation time, while they are not found in sources with longer estimated occultation times. Overall, our results show that occultations by BLR clouds  may be responsible for most of the observed X-ray spectral variability at energies higher than  2~keV, on time scales longer than a few ks.
\end{abstract}



\section {Introduction}
A circumnuclear, toroidal absorber is a key element in the structure of Active Galactic Nuclei (AGN), as implied by AGN Unified Models (Antonucci~1993, Urry \& Padovani 1995), and confirmed by a large set of observational data (e.g. Bianchi, Maiolino, Risaliti~2012 for a recent review). 
The composition, geometry and inner structure of this absorber has been long investigated, with evidence of a large spread of properties and dimensions, ranging from sub-pc to hundreds pc scales, and from largely homogeneous to clumpy structures. 
In this context, a key piece of information comes from measurements of X-ray absorption column densities (N$_H$): in almost all sources with multiple hard X-ray observations, the values of N$_H$ show high variability (Risaliti et al.~2002). This result implies a clumpy structure of the circumnuclear absorber, and an upper limit on its distance from the central source ranging from a fraction to several tens of parsecs, depending on the time separation  between the observations (and assuming Keplerian velocities for the obscuring clouds). 
In order to refine these measurements, it is necessary to investigate absorption variability at shorter time scales, with the aim of reducing the upper limits on the variation times (through campaigns of multiple observations of the same sources), or to directly measure them  (if the N$_H$ changes occur during single long observations).

Several such studies have been performed in the past few years, on a small number of sources: NGC~3227 (Lamer, Uttley \& McHardy~2003), NGC~1365 (Risaliti et al.~2007, 2009; Maiolino et al.~2010), NGC 4388 (Elvis et al.~2004), Mrk~766 (Risaliti et al.~2011), NGC~4151 (Puccetti et al.~2007), NGC~7582 (Bianchi et al.~2009; Piconcelli et al.~2007), UGC~4203~(Risaliti et al.~2010), NGC~4395 (Nardini \& Risaliti~2011);   SWIFT J2127.4+5654 (Sanfrutos et al.~2013). A review  of several  of these observations is presented in Risaliti~(2010). 

Most of these studies were performed following a two-step strategy: 

1) When a flux light curve shows strong  variability  and such strong variations are also observed in the  corresponding hardness ratio light curve, a change in the spectral shape must be present.  It is possible to interpret  the observed spectral variations  in two different ways: {\it i)} as due to changes in absorption properties, such as  
$N_H$ and covering factor, of clouds crossing the line of sight towards the X-ray source;  
{ \it ii)} as due to changes of the slope of the intrinsic continuum.

2) A complete spectral analysis is performed on each time interval defined in the first step, in order to  determine  whether   the observed variability can be explained by a change in the photon spectral index. In all the cases mentioned above, this analysis confirmed that the spectral variations were due to changes in the absorbing column density, occurring on  time scales from a few hours to a few days.

The physical consequences of these results are quite relevant in the context of AGN structure and physics. Assuming that the observed occultations are due to clouds orbiting the central black hole/X-ray source with Keplerian velocity, the main conclusions of the studies performed so far are the following:\\
-
The X-ray ``torus" is made of cloud-like structures
with the same density and distance from the center as  those of the broad line region (BLR)  emission line clouds (Bianchi, Maiolino \& Risaliti~2012). Therefore, in at least some sources, the X-ray absorber and the Broad Line region are one and the same (Maiolino et al.~2010).\\
- 
 The short transit time of these clouds across the X-ray sources, compared with the estimated black hole masses, is a direct observational proof that the X-ray source has a size of a few gravitational radii, as commonly assumed in disk+corona models (see for example Haardt \& Maraschi~1993).\\
- In a few cases with high S/N observations, we were able to investigate the structure of the obscuring clouds, inferring a "cometary" shape (Maiolino et al.~2010;  Risaliti et al.~2011).\\
- If X-ray occultations occur during an observation, they must be considered in the spectral analysis, even when the purpose of the study is not related to absorption-related issues. Alternatively, a wrong estimate of the other spectral parameters is possible, as demonstrated in the case of NGC~4395 (Nardini \& Risaliti~2011).

All these results have been obtained through a large observational and analytical effort on a small number of  ``ad-hoc" sources. The clearest example in this sense is NGC~1365, which is one of the most monitored AGNs, with  several observations with each of all the main X-ray observatories in the past few years, only because of its known high frequency of occultation events.

It is therefore fundamental to move from a ``one by one" analysis to a more general study of a statistically significant sample of AGN, in order (a) to estimate how common are X-ray occultations, and (b) whether the properties of the X-ray source and of the absorbing clouds inferred from previous studies are typical of AGNs.

This ambitious task requires the availability of a huge amount of high-quality X-ray data, which can be obtained only through archival observations, and a similarly large analysis work, which is probably beyond the capabilities of any single group. However, we believe we can obtain significant results with a simplified approach, based on a refined analysis of hardness ratio light curves (the first step in the strategy summarized above) and avoiding a complete time-resolved spectral analysis.
In this paper we present this new method, and the main results from a statistical analysis of its application to a large, well defined sample.
We will demonstrate that a careful time analysis can effectively discover occultations with durations of the same length, or shorter, than the observing time, and with column densities in the range 10$^{23}$-10$^{24}$~cm$^{-2}$. We will also show that these events are quite common among AGNs, implying that broad line clouds are in many cases the source of the observed X-ray absorption.

In Section~2 we define a sample of $\sim$40 X-ray bright AGNs with long X-ray observations, with suitable archival X-ray observations. In Section~3 we outline the method, which is based, and calibrated, on our past works on time-resolved spectroscopy of selected sources. In Section~4 we discuss the results, focusing on testing the ability of our analysis to discriminate between actual occultations and other  possible variability causes. In Section~5 we discuss the physical implications of our results, especially regarding the statistics of the occurrence of the occultations. Our conclusions are presented in Section~6, while a complete discussion of all the physical consequences of our results will be presented in a forthcoming, companion paper (Pietrini  et al.~2014, in preparation).  

\section{Observations: the sample}

Our aim is to analyze a representative, unbiased sample of local AGN. 
Our main requirements are the following.

 1) A high  X-ray flux, in order to detect possible spectral changes due to occultations with high statistical significance. In principle, our search may be extended to the whole X-ray domain, however, due to additional spectral complexities below 2~keV (soft excess, diffuse emission, warm absorbers), and to poor signal-to-noise above 10~keV, we will restrict our analysis in the 2-10~keV range. This implies that our search will be sensitive only to occultations by clouds with column density in the range 10$^{23}$-10$^{24}$~cm$^{-2}$  and that our sensitivity in disentangling spectral index variations and absorption variations is limited. We will further discuss this limitation in the next Sections.
 
2) A  ``long enough'' exposure time. The definition of a minimum acceptable duration is not straightforward, because both physical and technical considerations may be relevant. Operationally, we may define a minimum duration based on the source flux: the higher the  flux, the shorter the time needed to identify an occultation through spectral analysis. From a physical point of view, we may expect that the typical duration of an occultation depends on the size of the X-ray source and, therefore, on the black hole mass. Since we do not want to introduce biases based on our physical expectations, we did not consider the latter criterion in the composition of the sample; however its impact is discussed in detail in Section~5. 

3) High-sensitivity instruments, in order to obtain the highest possible count rate for each source.

 Based on the above requirements, we composed a sample of bright AGNs, consisting of all sources with  2-10~keV flux higher than 2$\times$10$^{-11}$~erg~s$^{-1}$~cm$^{-2}$, where the  X-ray flux was estimated extrapolating the measured  15-195~keV flux from the {\em Swift}/BAT 54~month catalogue (Cusumano et al.~ 2010). This choice is based on the completeness of such AGN sample in terms of {\em intrinsic} X-ray flux: absorption with column densities below $\sim$10$^{24}$~cm$^{-2}$ does not affect the observed flux in the 14-195~keV energy band (this is not the case for thicker absorbers: for this reason Compton-thick sources were excluded from our study, see below). In order to study the effects of absorption variability, we need bright sources in term of intrinsic, rather than observed, flux (i.e. in order to measure $N_H$ variations with small errors we need a high S/N at energies {\em above} the photoelectric cut off). For this reason, the Swift/BAT sample is preferred over other samples directly selected in the 2-10 keV band, such as the Piccinotti et al.~(1982) sample. Among these sources, we selected those with available observations made before January 2012,  by {\em XMM-Newton} and/or {\em Suzaku}, with durations of at least 80~ks and 50~ks, respectively.  The reason for a different choice of the observation duration for the two different  instruments is  the fact that the resulting useful  
 time interval is different in the two cases (see the end of Section 3.1 for an explanation).

Starting from this initial catalog, we made some "ad-hoc" corrections: (1) we excluded sources known to be reflection-dominated in the 2-10~keV range (such as NGC~4945, Mrk~ 3 or Circinus)    and other sources  for which {\em XMM} data are dominated by background flares or with calibration problems among the various  XIS instruments for
{\em Suzaku} observations (Centaurus~A,
QSO~B0241+62,  Mrk~6);  (2) we included NGC~4395 which, though slightly fainter than our flux limit, has been extensively searched for occultations (Nardini \& Risaliti~2011) and can be a useful comparison/calibration source for our analysis.

The choice of the instruments and of the observations duration is based on our past experience with single sources, and on the requirement of a final sample with a large enough size (42 sources,  85 observations) for a detailed statistical study, but still manageable for individual studies of each individual observation. 

The final sample is clearly not complete, but we believe it is truly representative of the AGN population, because all the 
adopted selection criteria are not related to the possible presence of occultations or, more in general, to the variability 
behaviour of these sources. The only possible exception may be the case of NGC~1365, which is over-represented in the sample, because the large number of available observations is primarily due to the the high frequency of detected X-ray occultations.

Moreover, several interesting cases remain out of our sample, such as the occultation studies based on campaigns of repeated short observation (e.g., UGC~4203, Risaliti et al.~2010), the extreme N$_H$ changes revealed by TOO observations triggered by long-term monitoring (e.g. NGC~4388, Elvis et al.~2004), and the long monitoring of the brightest sources made with lower effective area instruments (such as the eclipse discovered in a BeppoSAX observation of NGC~4151, Puccetti et al.~2007).
These cases can be discussed one-by-one, but are not included in our sample, to preserve its homogeneity.

The main properties of the sample, consisting of 32 {\em XMM-Newton} observations and 53 {\em Suzaku} observations of a total of 42 sources, are presented in Table~1.
It is apparent from Table 1 that our sample includes a wide range in source brightness and in central black hole masses as well as all
possible choices of source type ranging from ``pure'' Seyfert~1  to Seyfert~2.

\begin{table*}
  \centering 
  \begin{minipage}{150mm}
\caption{The sample of bright AGN with long, high S/N X-ray observations. ${\it (1)}$:  log of black hole mass in unit of $M_\odot$;   ${\it (2)}$: eclipsing time ($10^{3}$~s) 
computed as described in Sect.5,
 ${\it (3)}$: 2-10~keV  flux ($10^{-11}$~erg/cm$^2$/s),
 obtained extrapolating the {\em Swift}/BAT observed spectrum. ${\it (4)}$: Number of {\em  XMM-Newton} and {\em Suzaku} observations considered in the analysis 
 (more details on the single observations are presented in Table~2).  }

\begin{tabular}{lccccccccc} 
\hline

Name & redshift & Type & M$_{BH}$ &  $t_{ ecl} $& F(2-10) & ${\it XMM}$ &${\it Suzaku}$ &{ References} \footnote {
(1) Peterson et al. (2004), (2) Zhou et al. (2010), (3) Winter et al. (2009), (4) Winter et al. (2010), (5) Middleton, Done \& Schurch (2008),
(6) Beckmann et al. (2009), (7) Grandi, Malaguti \& Fiocchi (2006), (8) Rivers, Markowitz \& Rothschild (2011), (9) Denney et al. (2006),
(10) Whang \& Zhang (2007), (11) Malizia et al. (2008),  (12) Miniutti et al. (2009),
(13) Denney et al. (2010), (14) Kuo et al. (2011), (15) Grier et al. (2012), (16) Risaliti et al. (2009), (17) Peterson et al. (2005)
(18) Onken et al. (2007), (19) Bentz et al. (2006), (20)Guainazzi et al. (2010).
}\\
 & & &     $^{\it  (1)}$ &$^{\it (2)}$ & $^{\it (3)}$ & $^{\it (4)} $ & $^{\it (4)}$ & \\

\hline
NGC	 4151&	0.0033	&1.5&	7.66&120& 22.86 &	0	&2& 18,19\\
IC	4329A&	0.0160&	1	&8.3&1565& 18.32	&1&	0& 2\\
NGC	5506	&0.0062&	1.9	&6.7	& 112&16.70	&1&	2& 20\\
MCG	 5-23-16	& 0.0085&	2	&7.7& 428&14.73	&1&	1& 3 \\
NGC	4388	&0.0084&	2	&6.92	 & 152&11.58 &	0	&1& 14 \\
NGC	2110	&0.0078&	2&	8.3	&678& 11.31	&0&	1& 4,5\\
NGC	4507&	0.0118&	2&	8.4&1096&  10.63 &	0&	1& 3 \\
NGC	3783	&0.0097&	1.5	&7.5&132 &10.41	&2	&2 &1 \\
NGC	3227	&0.0039&	1.5	&6.9	&55& 7.06 &1&	1& 13 \\
MCG	08-11-011	&0.0205&	1.5&	8.1&1033&	6.87	&0&	1& 4,5,6\\
NGC	3516	&0.0088	&1.2&	7.5 & 147&6.76	&1	&2 & 13 \\
MCG	06-30-015&	0.0078&	1.2&	6.3&	63& 6.69 	&2&	3 & 2\\
3C	111&	0.0485&	1&	9.6	&11109& 6.64 &	1	&4 & 7 \\
MCG	02-58-022	&0.0469	&1.5&	8.4&2727& 6.44 &	0&	1& 8 \\
1H	2251-179&	0.0640&	1	&8.8	&5448& 6.03 &	0&	1 & 3 \\
Mrk	509	&0.0344&	1.2&	8.2&779&5.49 &	1&	1& 1,5 \\
3C	120	&0.0330&	1	&7.8	&457 &4.97 	&1&	2 & 15 \\
3C	382	&0.0579&	1&	9.2&7079 &4.54 &	0	&1 & 3,4 \\
IGR	 J21277+5656	&0.0147& NLS1	&7.18	&239& 4.44 	&1&	1 & 11,12\\
Mrk	110	&0.0353&	NLS1	&7.4&227	& 4.14 &	0&	1 & 1 \\
NGC	4593	&0.0090&	1	&7.0	&90& 4.10 &	0&	1 & 9 \\
NGC	7469	&0.0163&	1.2	&7.1	&155& 4.06	&2&	1 & 1 \\
Ark	120&	0.0327&	1	&8.2&685&	4.05 &	0&	1& 1 \\
NGC	7582	&0.0052&	2&	8.3&351&	 3.90	&1&	0 & 3 \\
NGC	1142&	0.0288&	2	&9.4	&4552& 3.84 &	0&	1 & 3 \\
NGC	4051	&0.0023&	1.5	& 6.2 &21& 3.75 &	0	&3& 13  \\
MCG	3-34-64	&0.0165&	2	&8.3	&915& 3.60 	&1&	0 & 3 \\
4U	0106-59	&0.0470	&1.2	&8.4&862&	3.50 &	1&	2 & 1 \\
NGC 	1365	&	0.0055&1.8&	6.3	&34&	3.36	&	2&	3& 16 \\
Mrk	79&	0.0222&	1.2	&7.7	&317& 3.36	&1	&1 & 1 \\
ESO	141-55&	0.0366&	1.2&	7.1	&377& 3.10 &	1&	0 & 10 \\
3C	445	&0.0564	&1.5	&8.3	&2119& 3.10	& 0 &	1 & 7 \\
EXO	055620-3820,2&	0.0340&	1.2&	8.4&1622&	3.07	&1&	0& 3 \\
NGC	 7314	&0.0048&	1&	6.0	&21& 2.99 &	0	&1 & 5 \\
NGC	 5548	&0.0172&	1.5&	7.6	&239& 2.92 	&1&	0 & 13 \\
PG	1501+106	&0.0364&	1.5	&8.15&503& 2.63 &0&	4 & 3,4 \\
Mrk 	766	&0.0129	&NLS1&	6.5	&83&	2.61	&	6&	2 & 6 \\
Mrk	279	&0.0305&	1	&7.5&258&	2.56 &	0&	1 & 1 \\
RHS	39	&0.0222	&1&	8.7&1368&  2.33 &1&	0 & 3,6 \\
2MASX	J04532576+0403416&	0.0296 &	2 &		&& 2.15 &	0&	1 & \\
NGC	1052	&0.0050&	2&	8.2	&240& 2.10 &	0&	1 & 6 \\
NGC	4395	&0.0011&	1.5&	5.56&2.6&	1.16		&	1	&0 & 17 \\
\hline
\end{tabular}
\end{minipage}

\end{table*}

\section{Our analysis}
\subsection{ Data reduction}

The data have been reduced and analyzed following a standard procedure, which we already described in other similar
works (e.g. Risaliti et al. 2009 for {\em XMM-Newton} data and  Maiolino et al. 2010 for {\em Suzaku} data).

For {\em XMM-Newton}, the light curves were extracted from the PN cleaned event files in the  circular region  including most of  the emission. For the bright sources of our sample
this radius results  $\sim  37$~arcsec. The background was extracted from a nearby region free of detectable sources. The high X-ray flux of our sources implies  that the background counts are in all cases negligible, and our analysis does not depend on the particular choice of the background region. 
The high source brightness is also the reason of the relatively large source extraction area: we find that such a radius optimizes the signal-to-noise in our light curves, by including the tails of the point spread function up to a $\sim$99\% encircled effective area.    

  For {\em XMM-Newton} data, the only non-completely straightforward 
aspect of the reduction is the choice of the
rejection level for the high-energy background flares. 
We have chosen a somewhat different approach depending on the observation mode.
For observations in ``small window" mode,
since our sources
are  bright, we have generally adopted  a threshold only slightly higher  than the conservative level suggested in
the {\em XMM-Newton}  data reduction guide for EPIC instrument, namely 1count/s. In the cases of ``full window" observations, we 
have chosen to allow for a relatively higher level, depending on the observation, with the aim of rejecting only the time intervals in which the 
background flares appear to saturate the instruments. In the subsequent analysis of the resulting light curves, we have checked 
that no apparently significant hardness ratio variation features were present right in those time intervals that a more conservative 
choice of the threshold would have rejected.
The selection of the appropriate time intervals has been
done analysing the  counts/s  light curves in the 10-13 keV energy band,  since   in this range, due to the strong cutoff in {\em XMM} effective area at $\sim$10~keV, a negligible contribution to the counts/s  is expected from the source and most of the detected  photons  originate from background proton flare events.

For {\em Suzaku} data,  light curves were obtained from a 2~arcmin circular region. The background was extracted from  clean regions in the same frames. All the available XIS	spectra have been used for each observation when the data were compatible.

Since our analysis is based on hardness ratio light curves, we need to have a relatively high number of counts in each time bin. For this reason, the light 
curves were generally grouped in 2000~s 
both for {\em XMM-Newton} and {\em Suzaku} data.

The hardness ratio ($HR$) light curves have been obtained from the ratio of the hard and soft curves. The exact energy range of these two curves is 2-4~keV and 5-10~keV, and has been carefully selected based on the procedure  described below. The error in the ratio has been estimated by applying a standard error propagation, assuming independent factors.  This may lead in some cases to an over-estimate of the actual uncertainties in the hardness ratio light curve, because the fluctuations of the hard and soft energy intervals at a given time are not always completely independent. In order to quantify this effect, we analyzed the dispersion with respect to a constant average value of the light curves where no variability was detected (a subsample of $\sim$15 unobscured sources). In almost  all cases the reduced $\chi^2$ is higher than 0.8, with two exceptions with values around 0.5. An F-test showed that these values are in most objects compatible with purely statistical errors, with a zero-hypothesis threshold of 5\%. When the probability of purely statistical errors was lower than 5\%, we estimated the average fraction by which we should decrease the measured errors in order to obtain a 5\% probability. These fractions are negligible in all cases, except for the two objects mentioned above, where they are of the order of 20-25\%. We conclude that the hardness ratio errors calculated assuming independent errors in the two energy intervals are a good approximation of the true statistical errors. If in a few cases systematic effects added a small contribution to the errors, this would imply an under-estimate of the significance of possible $HR$ variations, but not the inclusion of any spurious occultation event. 

 For each source, reduced and analysed data are indicated in Table 2 and identified by the observation date
and by the observing instrument.  The effective duration of the observation, $\Delta t_{eff}$ (expressed in ks), is also indicated.
 By ``effective'' duration we imply 
the total time extent over which the light curve is actually reliable, that is 
excluding those intervals in which background ``proton'' flares may affect the resulting counts, as far as {\em XMM-Newton} light curves are concerned.
 On the other hand, for {\it Suzaku} observations, the time interval over which a source is actually observed is in general significantly longer than
the reported exposure time (mainly because of Earth occultations), therefore our ``effective'' duration represents the total time interval over which we can analyze the light curve behavior.

 \subsection{Choice of the hardness ratio energy bands}

 The choice of the optimal energy intervals for the computation of the hardness ratio is not straightforward, because the measured counts in each energy band depend on the energy-dependent effective area. Therefore, we need to 
 make the convolution of several different models (corresponding to several different absorbing conditions) with the instrumental response and thus calculate
 the expected counts/s, in order to  identify which energy bands are most appropriate for our analysis, that is maximise  the observable effects of absorbing column density differences.
Obviously this choice also depends on which range of column density we are most interested in. 
In order to simplify our analysis, we decided to neglect all the information below 2~keV. In this way, we loose sensitivity to low-column density (N$_H$$<$10$^{22}$~cm$^{-2}$) variations, but we also avoid the complications related to the possible presence of warm absorbers. In principle, this is not an absolutely needed simplification: complete spectral fits can easily distinguish between warm and neutral absorbers. However, this requires some extra effort in the interpretation of the light curves, and is left for future extensions of our work.

Within the 2-10~keV band, 
we have computed photon fluxes for different energy ranges, namely $F$(2-4~keV),  $F$(3-5~keV), $F$(5-10~keV),  $F$(6-10~keV),
assuming an emitted spectrum consisting of an absorbed power-law model, with photon
index $\gamma$,  where $N_H$ is the column density of the absorbing gas of the cloud crossing the line of sight and $C_F$ is the intervening absorbing gas covering factor. 
We 
used the  XSPEC  package to 
   calculate the expected counts/s in each energy range, convolving our models with the instrumental response matrices for 
  both {\em  XMM } EPIC and {\em Suzaku} XIS instruments, and to derive, from these  ``synthetic" fluxes, the expected hardness ratio.
This procedure has been performed  for different values of $N_H$ and of the covering factor $C_{F}$ due to the cloud passage, assuming  $\gamma =2$ as  a representative value  for the photon index. 
 Four different hardness ratio combinations, namely $HR_1 (N_H, C_F)= F$(5-10~keV)/$F$(2-4~keV),
$HR_2( N_H, C_F)= F$(6-10~keV)/$F$(2-4~keV), $HR_3( N_H, C_F)= F$(5-10~keV)/$F$(3-5~keV), $HR_4( N_H, C_F)= F$(6-10~keV)/$F$(3-5~keV),  have been computed  together with their respective errors, $\sigma_i$ .\\
In order to quantify which one, among the  hardness ratio combinations  $HR_i$ defined above, undergoes the  largest variation owing to an obscuring  cloud crossing the line of sight, we have compared each combination with the corresponding ``unperturbed" hardness ratio, $HR_i(N_H=0)$, evaluating the significance of the difference by means of the quantity
$$\Delta_i = {HR_i (N_H, C_F)-HR_i(0,0) \over [\sigma^2_i(N_H, C_F)+\sigma^2_i (0,0)]^{1/2}}$$
defined for each combination  ($i=1, 2, 3, 4$) and  for different values of $N_H$ and covering factor $C_F$.   We note that in our models we did not include the iron line since, for the column density values $N_H <  7\times 10^{23}$ cm$^{-2}$ that  we are dealing with, its contribution to the hard band counts is negligible.
  The comparison among the obtained $\Delta_i$ values for $N_H$  in the range  $1\times10^{23}< N_H <  7\times 10^{23}$ cm$^{-2}$
shows that  the  $\Delta_i  $  behaviour  obviously depends on the choice of the soft energy range in the hardness ratio denominator. We obtain two different
``groups'' which differ essentially in their trend at  low values of  the optical depth, i.e. for $N_H <  3\times 10^{23}$ cm$^{-2}$. Hardness ratios
computed  with   the 2-4 keV energy band  show larger $\Delta_i$ values with respect to those evaluated using the 3-5~keV band, owing to the  energy position of the absorption cutoff.  Within each ``group" we have 
 $\Delta_1 \ga \Delta_2$ and  $\Delta_3 \ga  \Delta_4 $, indicating that the range 5-10 keV can be  a better choice for the hard band. This result is due to the larger
 number of counts  when the energy band 5-10 keV is used, due to the sharp decrease of  the instrument effective area (of both {\em XMM } EPIC and {\em Suzaku}  XIS) at energies higher than 
 6~keV.   As a consequence, the choice   $HR= F$(5-10~keV)/$F$(2-4~keV)  maximizes  the hardness ratio variation for $N_H$ values in the range $ 1\times10^{23}< N_H <  3\times 10^{23}$ cm$^{-2}$;  this result holds also for the case of
 a partial covering of the source ($C_F$ = 0.5). Therefore, we conclude that the choice $HR \equiv HR_1$  is the most appropriate to our purposes. The same is true in the case of Seyfert 2 sources for which a constant absorption $N_H=1\times 10^{23}$ cm$^{-2}$ is assumed in addition to the variable cloud absorption.\\

\subsection{A systematic search: our method of investigation}

 For each of the sources in our sample, we performed a quantitative analysis of the hardness ratio  light curves, consisting of the following steps.

\begin{figure}
\includegraphics[width=8.5cm]{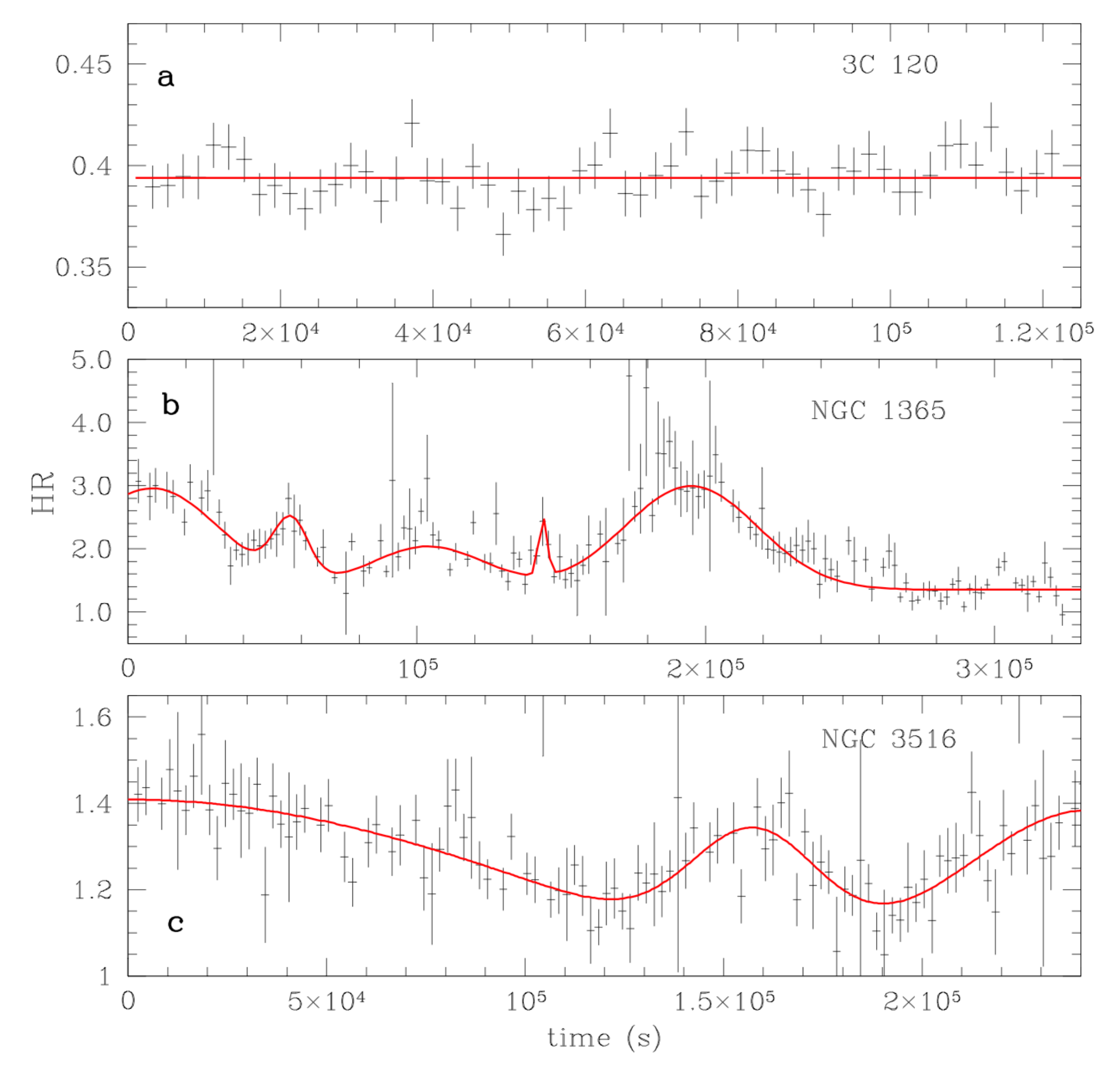}
\caption{  Three different examples of hardness ratio light curves and of the computed data fit (red continuous curves): ~a) the 08/26/03 {\em XMM-Newton} observation of 3C~120;
       b)  the 01/21/08 {\em Suzaku} observation of NGC~1365;  c)  the 10/12/05 {\em Suzaku} observation of NGC~3516.
}
   \label{examples}
\end{figure}

 From the obtained  $HR$ light curves, it is apparent that there are sources for which  the hardness ratio
is largely  constant in time, while for other sources there are  time intervals, within the observation length, in which the  $HR$ value undergoes
slight or  significant variations.  In order to point out    and to quantify this variability
 the following procedure has been adopted.
First of all,  each $HR$ light curve has been fitted with a constant, $K_0$, and the value of the 
corresponding reduced ``chi-square" $\chi^2/d.o.f. \equiv \chi^2_{r}$ has been computed;  we refer to the reduced ``chi-square" for the case of the initial fit with a constant only as  $(\chi^2)_0/(d.o.f.)_0 \equiv \chi^2_{r0}$. The values of these three quantities [$ \chi^2_{r0}$, $(d.o.f.)_0$, $K_0$] are listed in Table 2.
There are cases in which the $ \chi^2_{r0}$ value is close to unity,
 thus ensuring that the  fit with a constant value gives  a good representation of the $HR$ light curve.
 These are the cases in which no candidate eclipsing event can be identified from the observation and 
 no further analysis is to be performed.  As an  example, the {\em XMM-Newton}
  observation  of 3C~120 performed on 08/26/2003  is presented in Fig.~\ref{examples}\,a, where the
 hardness ratio light curve is shown, together with the fit with a constant   [$(\chi^2)_0/58=\chi^2_{r0}=0.98$].
 
On the other hand,  there are several cases for which  the opposite conclusion holds. Although the true goodness of a fit depends {\it both}  on the
 $ \chi^2$ value {\it and} on the number of degrees of freedom ($d.o.f.$), we {\it operationally} choose to  define
the value $(\chi^2)_0/(d.o.f.)_0= 1.1$ to be the  limit between the two classes of cases mentioned above, as a simple way to {\it conservatively} select observations deserving further study. For the cases with   $\chi^2_{r0}> 1.1$ a better
fit is seeked for. In order to perform a new fit, we define a general form for more complex models for the $HR$ light curve,  described by the following  functional form  $G_N(t)$:
\begin{equation}
G_N(t)=K_N+ \sum_{n=1} ^N A_n e^{-{1 \over 2}({t-t_n \over w_n})^2}.
\end{equation}
In the above expression  we suppose to represent the trial function as a constant plus a number N of different gaussian components.
 For each n-th gaussian component,  $t_n$ defines the gaussian  center, i.e. the time at which the $HR$  reaches
its  local  maximum value;   $A_n$ is the normalization of the n-th gaussian component and  is positive definite, since an occultation is bound to produce an increase in the hardness ratio value  above the  constant level, $K_N$,  and  $w_n$ is related to the width of the same n-th gaussian component.

Hence,  for light curves such that the  fit with a constant function (i.e. , $N=0$, $G_N(t)=G_0(t)$) results in  $\chi^2_{r0} > 1.1$, a new fit is performed  using ${\it G_1}(t)$ with $N=1$, that is testing a model composed of a constant plus one gaussian component, as a first step to improve the description of the light curve, and a new value of  $ \chi^2=  \chi^2_1$ is obtained. If this new fit is still not ``satisfactory'', we proceed further on by adding another gaussian component to the model function
($G_2(t)$, $N=2$) and re-fitting the light curve. This step-by-step procedure for  improving the model  descriptive capability by adding one gaussian component at a time should go on until a ``satisfactory'' fit is obtained with a model function $G_N(t)$ including  a number $N$ of gaussian components.
The procedure leads to a meaningful result only if   we define  a quantitative measure for a fit to be more ``satisfactory''  than the previous, less complex, one. Therefore, in order to  quantify the  improvement  of each new step fit 
with respect to the previous one   and hence to estimate the significance of each newly added gaussian component, we have
applied the F-test.
The F-test  calculates the  null probability value,  ${\it F}_n$,  on whose basis one can reject the simpler model (as worse) and maintain the more complex one (as a better description of the data) or not. 
We accept the  more complicated model as the  best one if the ${\it F}_n$ value is less than 0.05. 
  Following this method,   for  each step of our procedure, we use the F-test to compare the fit with a function including $n$ gaussian components plus a constant with the previous step (the ``simpler'' one), for which the describing model is defined by $(n-1)$ gaussian components plus a constant. This way, the null probability value,  ${\it F}_n$, that we obtain is indeed a measure of the significance of the additional  $n-$th gaussian component.  To exemplify, comparing the $G_0(t)= K_0$ model, representing the light curve with a simple constant, with the next fitting step, including one gaussian component and described by 
$G_1(t)=K_1+  A_1 e^{-{1 \over 2}({t-t_1 \over w_1})^2}$, the null probability value ${\it F}_1$ gives the significance of the $HR$ time variation that we attempt to describe with the gaussian component defined by the parameters $\{A_1, t_1, w_1\}$, given by the fit; the same will occur at each further step, allowing us to associate a significance estimate $F_n$ to all the gaussian components  $\{A_n, t_n, w_n\}$ (with $n= 1...N$) that compose our final fit (attained when 
${\it F_n}$ becomes larger than 0.05).

 An  example of  this case is presented in Fig.~\ref{examples}\,b  where  the hardness ratio of the 01/21/2008 {\em Suzaku} observation of NGC~1365 and the derived fit with  5 gaussian components are shown.
  As it is apparent, in this case $HR$ is a variable
function of time, but during the last third of the observation interval its value becomes almost constant.  This fact is important, since the fit can easily determine
the constant $HR$ level to which  the gaussian components induced by occultations  add up.

The values of the quantities ($A_n,  F_n $), i.e. the peak value of the n-th gaussian component and its significance, are reported in Table
2  for those components having ${\it F}_n< 0.05$.  As described before, we follow this procedure until ${\it F_n}$ becomes greater than 0.05. However, as we shall  discuss in the next Sections, we will choose a much more conservative significance threshold to identify the ``safe'' occultation events.
  
We note that our choice of gaussian functions is arbitrary, but quite effective. Since the shape of the $HR$ light curves is not predictable, our aim was to reproduce the most significant variations with simple functions, defined just by  some measure of height and width. A triangular function, or similar shapes, would also be acceptable. However, the choice of the exact shape of the function does not affect our analysis and our results.

There are some cases  in which the above procedure fails because the light curve oscillations
are very close in time  and hence it is almost impossible to disentangle  the underneath constant level
from the gaussian wing contributions. In these cases the only way to proceed is to fit the light curve with an appropriate
$ G_N(t)$  function and then eliminate every gaussian curve in turn, comparing the [$N-1$]-gaussian components fit with the complete one to
find the significance of the suppressed hump. 
 An  $HR$ light curve example in which  the  ``unperturbed" constant value of $HR$ is not immediately identifiable at first glance is the light curve shown in Fig.~\ref{examples}\,c, together with the corresponding best fit we obtained, representing the 10/12/2005 {\it Suzaku} observation of NGC~3516. In this case, within the observation duration no  clearcut constant $HR$ time interval is present and a definition of the constant level underneath the variation components is possible only through the fitting process: our best fit defines 3 gaussian components that can be easily recognized in the figure.  For this specific case our result is confirmed by the spectral analysis (Ursini et al.~2014)
from which the authors conclude that the three components shown in Fig.~\ref{examples}\,c  are originated by the interposition of an obscuring cloud and
 cannot be due to a change in the emitting source spectral index. This fact confirms the reliability of our procedure also in the cases in which the various components are
 rather mixed.
 
 We point out that in cases like the  present one, we have also attempted a fit allowing for negative values of $A_n$. Not always 
such attempts produce   good fits, but in some cases, like for the observation  of  NGC~3516 discussed above, the fit in which the $HR$ light curve
is described by a constant level ($K_N=0.65$)  with additional positive gaussian curves ($A_n> 0$) and  the one  described  by a constant level ($K_N=1.38$) 
 with additional negative    gaussian curves  ($A_n< 0$) 
are comparable, i.e. they result in almost equal values of the reduced $\chi^2$.   However,  the  case with negative gaussian curves  demands a high value
for the $HR$ constant level ($K_N=1.38$) which could be reproduced, in a type 1.2 source like NGC~3516, only by a very low   intrinsic  photon index. Such   a flat
spectrum in an unabsorbed source is not consistent with the observed properties of this type of sources.
This conclusion also holds 
for the other few sources for which a fit with negative gaussian curves was possible.

As described above, fits   with ``high''  initial reduced $\chi^2$ ($\chi_{r0}^2>1.1$ )  are  followed by a further fit and the significance of
adding a new gaussian to the fit is computed.  However, for some cases this procedure 
does not give improving results, since the  F-test shows  that the simpler model, i.e. the fit with a constant function, is a better representation of the data.
 We believe that for these $HR$ light curves  the high
$\chi_{r0}^2$ value is most probably a consequence of  a 
``degraded''  quality of the data, rather than being due to the actual presence of variations. 

Looking at Table~2, it is apparent that  there are many sources for which the $HR$ fit with a constant function is
a good one, thus indicating that the light curve is  basically constant in time. On the other hand there are
 several sources for which one or more observations show a better fit of the light curve  when a time dependent trial function, i.e. ${\it G_N(t)}$, is used.  These cases are good candidates for occultations.

Besides the statistical significance of $HR$ variations, a second relevant  indicator obtained by our fits is    the ratio between the peak height of the $n$-th gaussian curve, and the underneath constant value ($A_n /K_N $ in our model representation) which  is a measure of the relative  variation of the hardness ratio, $\Delta(HR)/HR$.
The importance of this parameter will be clear from the analysis in the next Section, where we discuss the possible alternatives to occultations to reproduce 
the observed $HR$ variations. 
Qualitatively, events with relatively low values of  $\Delta(HR)/HR$, but with high statistical significance, may be found in light curves with particularly  high S/N. In these cases, the spectral changes would be small despite the high statistical significance of the  event. Therefore, the parameter $\Delta(HR)/HR$ is needed to better chacterize the $HR$ variation events.

\begin{figure}
\includegraphics[width=8.5cm]{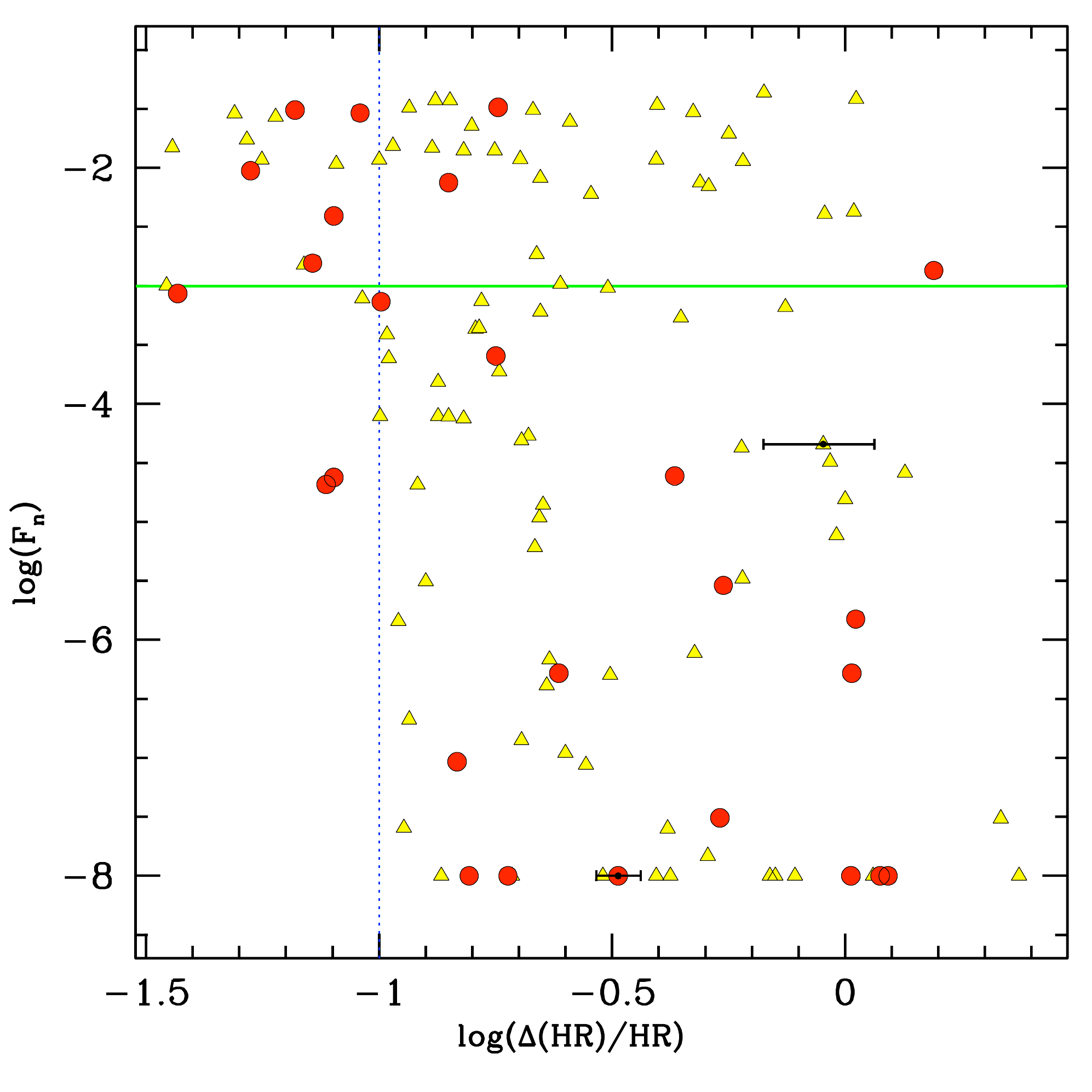}
\caption{ Significance value, $F_n$, for each hardness ratio variation  fitted with a gaussian (see text for details), as a function of the corresponding  relative peak height, $\Delta(HR)/HR$. Each variation is plotted as a triangle. Big circles represent the most significant event for each source in our sample. Note that for peaks with  significance $F_n< 10^{-8}$ the value $F_n= 10^{-8}$ is shown in the figure.
  The 16  sources for which no significant ($F_n> 0.05$) event has been detected are not shown in the diagram. 
  Representative error bars for  $\Delta(HR)/HR$ are reported for a couple of events shown. }
   \label{diagnostic}
\end{figure}

The results of our analysis are shown in Fig.~\ref{diagnostic}, where for each event the probability, $F_n$, that it is due to a statistical fluctuation 
is plotted against its $\Delta(HR)/HR$ ratio.  Two representative error bars  for the quantity $\Delta(HR)/HR$,  shown in the figure,  illustrate the error trend resulting from our analysis, i.e. the fact  that smaller errors are associated to lower $F_n$ values.
It is  also apparent that significant statistical ($F_n \leq 0.001$) $HR$ variations are  typically  associated with  $\Delta (HR)/HR >0.1$,  while this limit 
does not hold for  $HR$ humps associated to $F_n$ values $>0.001$.  The implications of this result
will be analysed in Section 5.

%
%
 
%
%

\section{Interpretation of the  analysis}

The possible interpretations of humps in $HR$ light curves as due to occultation by interposing clouds must be confirmed through a complete spectral analysis. For some sources in our sample this analysis  has already been done. For example, as reported above,
 Risaliti et al. (2011) interpret as eclipses three strong variations in the hardness ratio light curve of Mrk 766  during {\em XMM} observations  of 23 and 25 May 2005. 
  The same  interpretation has been confirmed by detailed analyses  of  $HR$ variations observed in other sources (e.g. NGC~4151, Puccetti et al.~2007; NGC~4395, Nardini \& Risaliti~2011; NGC~1365, Risaliti et al.~2007, 2009).

Here we discuss the main possible sources of  $HR$ variability, and we test our interpretation on two light curves for which we already performed a complete spectral analysis.
All the results shown in this Section are obtained through  the convolution of  the theoretical models  with the {\em XMM}-EPIC/PN response matrix,  i.e. accounting for  the effects of the energy-dependence of the effective area.  Since the result depends on the specific instrument, when needed,  the convolution is performed separately for the {\em XMM-Newton}-EPIC/PN, and the {\em Suzaku}/XIS instruments. In the following, we  shortly call the quantities resulting from  this procedure  as ``synthetic".

\begin{figure}
 \includegraphics[width=8.5cm]{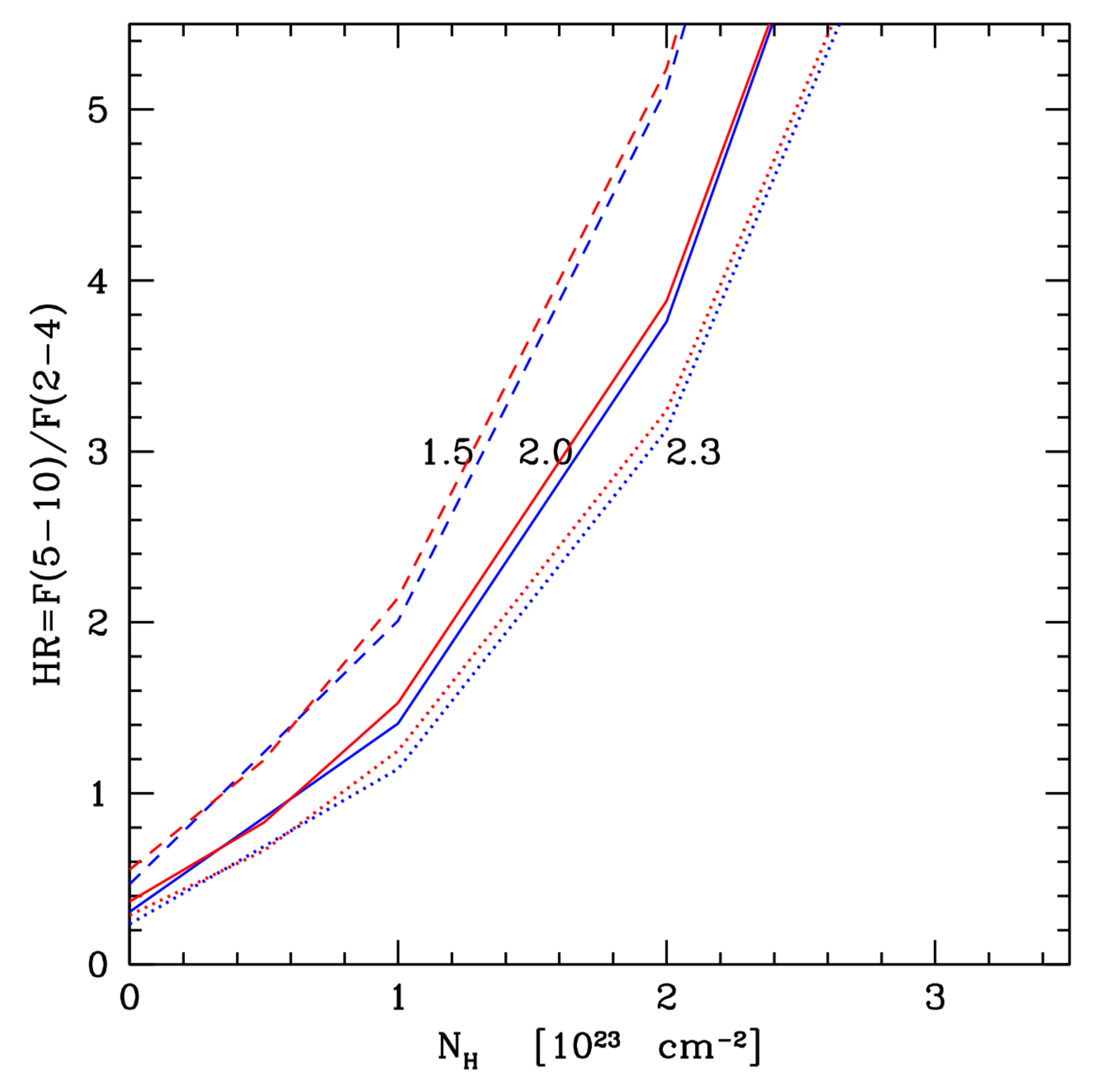}
   \caption{Synthetic hardness ratios  as functions of the column density $N_H$ for different values of the spectral index $\gamma$ and for $C_F=1$ and $R=0$. For each $\gamma$,  convolutions with the {\em XMM} (blue lines) and {\em Suzaku} (red lines) response matrices are shown.}
   \label{HR}
\end{figure}

 A complete analysis of the hardness ratio  variability implies the analysis of its dependence  on $\gamma$, $N_H$, the cloud covering factor, $C_F$,  and the ratio between the normalization of the reflected to the direct power law
components,  $R$, that is the evaluation of $HR\equiv HR(\gamma, N_H, C_F, R$).       As a first insight, the 
 general behaviour of $HR$ for different column densities and photon indexes is shown in Fig.~\ref{HR}.
The visual analysis of Fig.~\ref{HR} reveals several significant  properties of our indicator:\\
- The value for unobscured sources is of the order of 0.4, with little dependence on the photon index.\\
- The dependence on the photon index is weaker than that on the column density: the plotted lines refer to the whole observed range of photon indexes in nearby AGNs, nevertheless their distance in the $HR-N_H$ plane is not large, compared with the range of $HR$ corresponding to variations of column density of the order of 10$^{23}$~cm$^{-2}$. 

To better quantify the different origins of  $HR$ variability and  in order to compare  our ``synthetic" results
with   those of the data analysis shown in Fig~\ref{diagnostic}, we  
 computed for the hardness ratio function,  $HR\equiv HR(\gamma, N_H, C_F, R$),
its  fractional  variability, $\Delta(HR)/HR$, with respect to each of  the  independent variables in turn,  
keeping all other variables constant.
In the next Subsections we investigate these issues in more detail.

\subsection{Effects of spectral index changes}

The obvious  objection to the interpretation of humps present in the $HR$ light curves in terms of interposing clouds is
 that a spectral change, namely a decrease in the spectral index value, $\gamma$,  of the X-ray emitting source, would produce  the same effect.
In this framework, the humps
in the $HR$ light curves could be the result of an episodic hardening of the emitted spectrum.

In the hypothesis that  $HR$ changes are only due to spectral index changes, we tested
several cases maintaining $C_F=1$ and $R=0$. We noticed that for any given variation of photon index $\gamma$, we observe the highest  fractional change  $\Delta(HR)/HR$ in unobscured sources. Therefore, we concentrated on spectra with $N_H$=0, in order to explore the highest possible variations and hence we evaluated
\begin{equation}
$$ \big [{\Delta (HR) \over HR}\big ]_{\gamma}= {HR(\gamma - \Delta \gamma, 0,1,0)-HR(\gamma, 0,1,0) \over HR(\gamma, 0,1,0)}$$
\end{equation}
 where $HR(\gamma, 0,1,0) \equiv HR(\gamma, N_H= 0, C_F=1, R=0)$.

\begin{figure} 
\includegraphics[width=8.5cm]{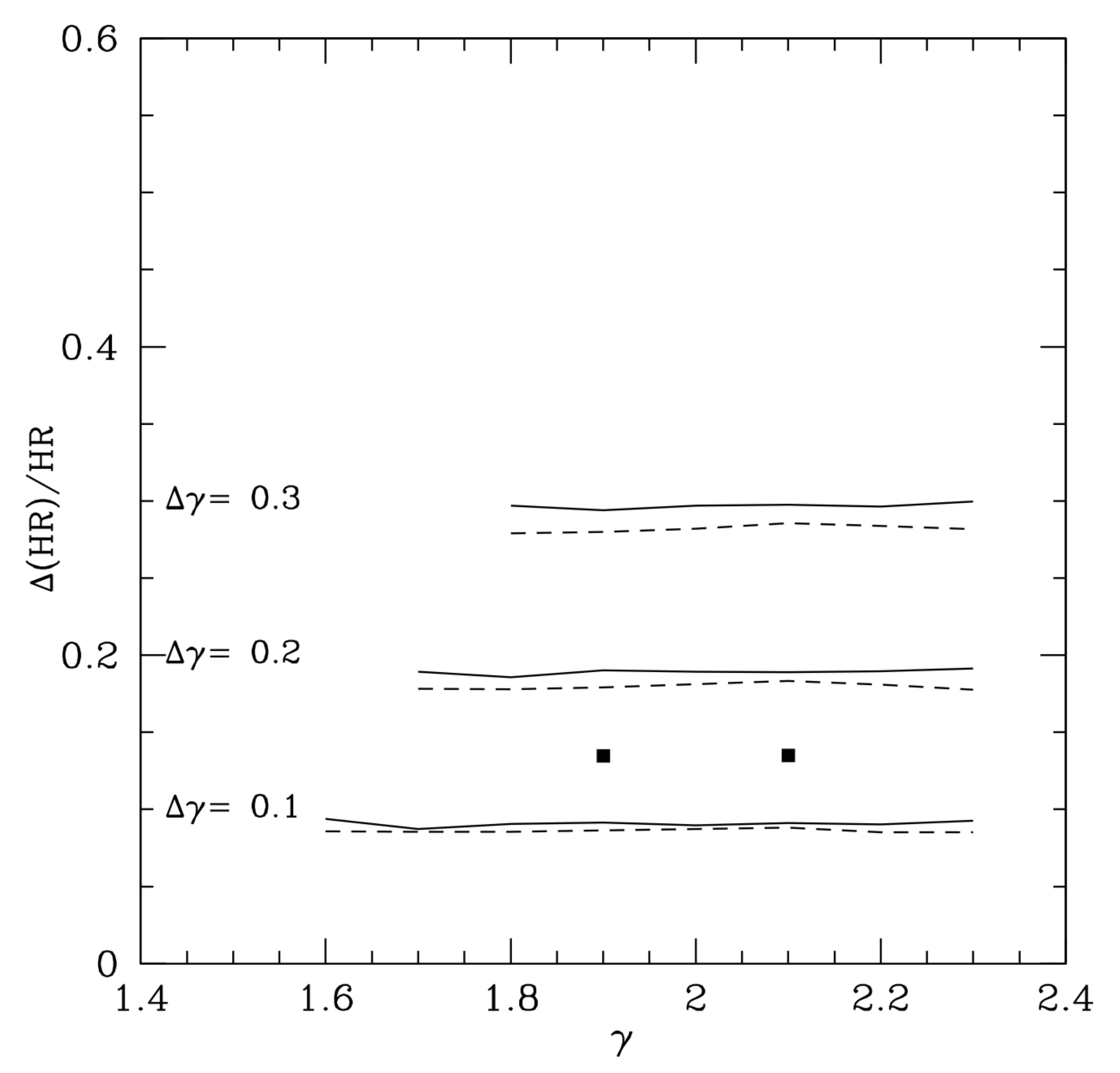} 
\caption{Upper limits of $[\Delta (HR)/HR]_{\gamma}$  values obtained from synthetic {\em XMM} (continuous lines) and {\em Suzaku} (dashed lines) spectra. Curves are labelled with the corresponding  $\Delta$$\gamma$. The two square points refer to the case $\Delta$$\gamma$=0.2 for {\em XMM} synthetic model  and show the effects of a  constant non zero $N_H$, i.e.
$N_H =2\times10^{23}$ cm$^{-2}$. }
\label{Dgamma}
\end{figure}

In Fig.~\ref{Dgamma} we present 
 synthetic  $\Delta(HR)/HR$ values obtained  starting from different values of $\gamma$, and corresponding to three variations. Since we assume a positive $\Delta \gamma$,   a decrease of the
spectral index value results in an increase  in $HR$, as Fig.~\ref{Dgamma}  shows.  In addition, these computations show two interesting facts: 1) the effect of $\gamma$ variability is nearly the same for {\em XMM-Newton} and {\em Suzaku} spectra. This is evident also in Fig.~\ref{HR} and  it is due to the similarity of the dependence of the effective areas on energy for the two instruments. As a consequence, from now on we  show only the results for {\em XMM-Newton}. Anyway, in all the cases discussed here we 
 computed the synthetic quantities for both the observatories; 2) the effect of a variation of $\gamma$ is almost independent of the starting value of $\gamma$. This effect is also due to the smooth behaviour of the effective area as a function of energy. Moreover, the measured $[\Delta (HR)/HR]_{\gamma}$ scales  almost linearly with $\Delta$$\gamma$. As a consequence, we can approximate $[\Delta (HR)/HR]_{\gamma} \sim \Delta \gamma$ for every value of $\gamma$. 

The figure also shows two square points obtained from {\em XMM} synthetic models  for  $\Delta(HR)/HR$  in the case of a constant non zero $N_H$, i.e. $N_H =2\times10^{23} $cm$^{-2}$, taken as representative  value for Type~2 sources. These  two points  refer to the case of  $\Delta \gamma=0.2 $ and therefore confirm what previously  noticed, {\it i.e.} that the curves computed for $N_H =0~$ cm$^{-2}$ and relative to Type~1 sources,  those drawn in Fig.~\ref{Dgamma},  represent
an upper limit to changes in $HR$ curves due to spectral index variations.

    \noindent  
 We have reported in Fig. \ref{Dgamma}  the cases in which $\Delta \gamma \leq 0.3$. However, we do not expect any $\Delta \gamma \geq 0.1$ during individual observations, based on the two following arguments.\\
- Sobolewska \& Papadakis~(2009),  and Papadakis et al.~(2009), 
 report a $\gamma$-intrinsic flux correlation in a sample of bright AGN (most of their sources are included in our sample), based on hundreds of snapshot Rossi-XTE observations over a period of more than 10 years. This correlation is obtained from a power law fit of each spectrum, so the physical origin of the correlation is not a-priori clear (for example, variable absorption could play a role, with a flatter spectrum and a lower flux associated to more absorbed states). However, even assuming an intrinsic $\gamma$-flux correlation, in most cases the slope of the dependence is flat enough to imply a photon index variation $\Delta\gamma<0.1$ in the flux variability range observed in single observations (an exception is NGC 4051, which is known to show large flux and photon index variability, associated to either variable absorption or changes in the reflection fraction  (Ponti et al.~2006)). \\
- In order to check this important point, we performed a complete spectral analysis  on a subsample of five bright unobscured AGNs with significant $HR$ variability (those with no spectral variability 
obviously do not need any check to confirm the constancy of $\gamma$). In this study, presented in the second paper of this series (Ursini et al.~2014) we fully confirm that all the cases with $\Delta(HR)/HR>0.1$ are best reproduced through occultation events, while the variability in the only source in the sample with $ \Delta(HR)/HR <0.1$ can be also explained through intrinsic variability of $\gamma$. 
 
\subsection{Effects of large flux variations}

\begin{figure} 
 \includegraphics[width=8.5cm]{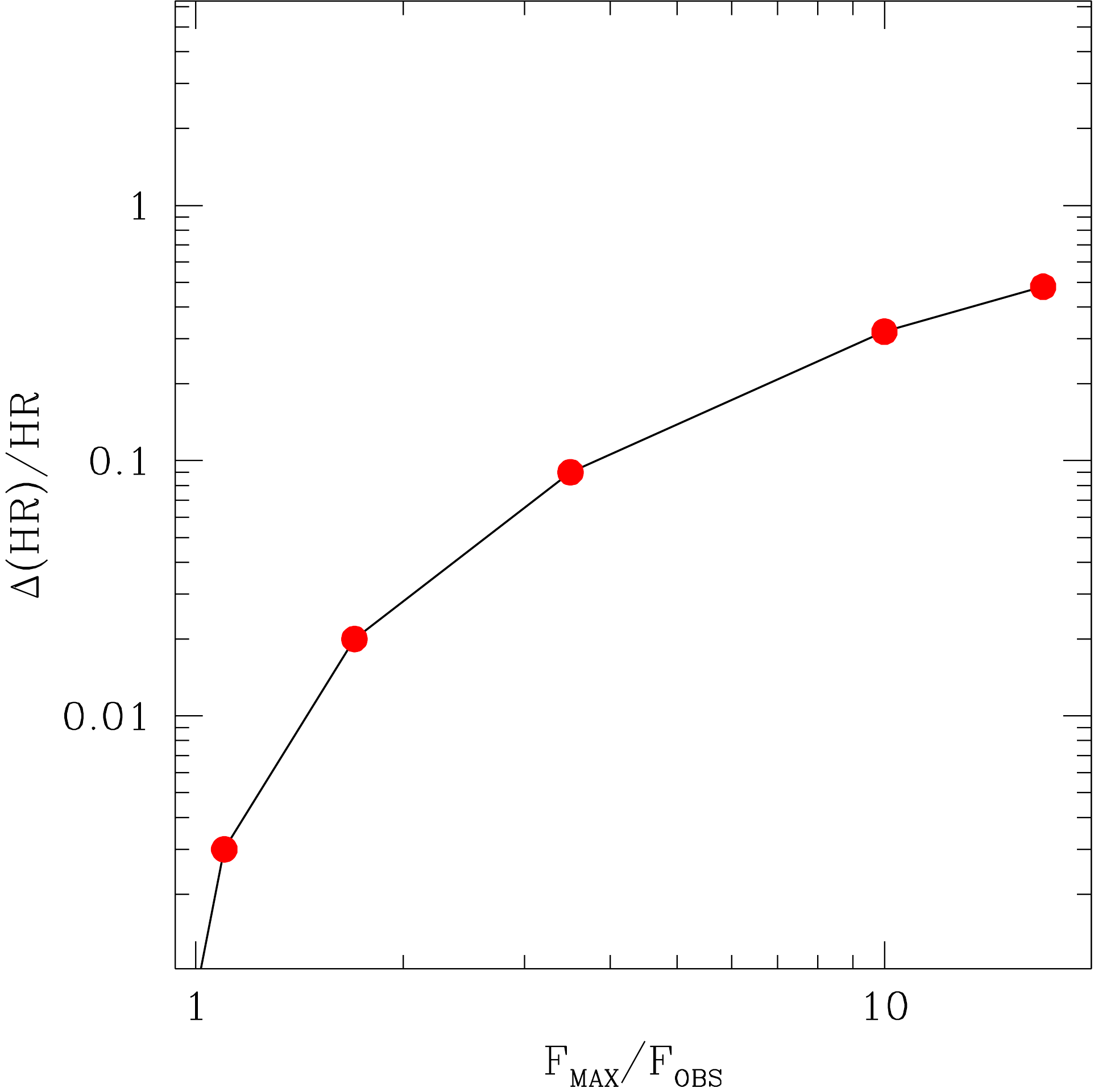}
 \caption{Computed  $\Delta (HR)/HR$ corresponding to 2-10~keV intrinsic flux variations. To compute the synthetic quantity,
 we assume that the intrinsic flux drops from $F_{MAX}$ to $F_{OBS}$, while the reflection component remains at a constant value, corresponding to the reflection of $F_{MAX}$ by a Compton-thick reflector, covering 2/3 of the solid angle.
}
   \label{refl}
\end{figure}

A possible source of hardness ratio variability may be extreme {\em flux} variability. In most nearby AGN, a cold reflection component is observed, with a 2-10~keV flux of a few percent of that of the intrinsic emission. The total observed spectrum is therefore obtained adding these two components. If the total spectrum is fitted with a single power law, the measured best fit photon index is close, within a few percent, to that of the intrinsic power law, the contribution of the reflection being small. However, if the reflector is far enough from the central source, a strong decrease of the intrinsic flux may propagate to the reflection components in longer times than the intrinsic variability time scales. In this scenario, it would be possible to observe quite large reflection/intrinsic emission ratios. Since the reflection component is much harder than the intrinsic power law in the 2-10~keV range (e.g. Magdziarz \& Zdziarski~1995,  Murphy \&Yaqoob~2009) we would then observe a significant change of the observed $HR$. This is indeed one of the explanations suggested by Sobolewska \& Papadakis~(2009) to explain the $\gamma$-flux correlation over large flux variations. 
 In order to explore this possibility we computed the synthetic quantity $\Delta(HR)/HR$
assuming a configuration in which we have 
an intrinsic power law continuum with constant (fixed) photon index, $\gamma$=2, and no intrinsic absorption, together with a 
Compton-thick reflector covering 2/3 of the solid angle as seen from the central source, originating the reflected emission component.
Technically, we simulated this scenario through a PEXRAV model (Magdziarz \& Zdziarski~1995) with a ratio R between the normalizations of the reflected and intrinsic components R=1.35.  Since the actual reflectors are on average less efficient than the one assumed here, the results of our procedure can be considered upper limits of the observable $HR$ variations. 
Within this framework, we suppose that, due to intrinsic source variability, the intrinsic flux decreases from a ``high'' value $F_{MAX}$ to  a value $F_{OBS}$ and we assume 
that within the intrinsic variability  time-scale (i.e., the lapse of time in which the intrinsic flux value drops) the reflection component stays constant, because the propagation time for the intrinsic flux variation to reach the reflector distance is longer than the intrinsic variability time-scale.  If this is the case,   the synthetic $HR$ variations are
induced by the different {\it relative}
contribution of the reflected component in the two conditions for the intrinsic flux.
The results are shown in Fig.~\ref{refl}.

It is clear that the level of X-ray  flux variability usually observed in AGNs, on time scales of hours-days, is not enough to produce high changes of $HR$.

 \subsection{Effects of $N_H$ variations}

The same procedure described in Section~4.1 can be applied to the case of interposing clouds with different covering factor, $C_F$,  but still with $R=0$.

\begin{figure}
   \includegraphics[width=8.5cm]{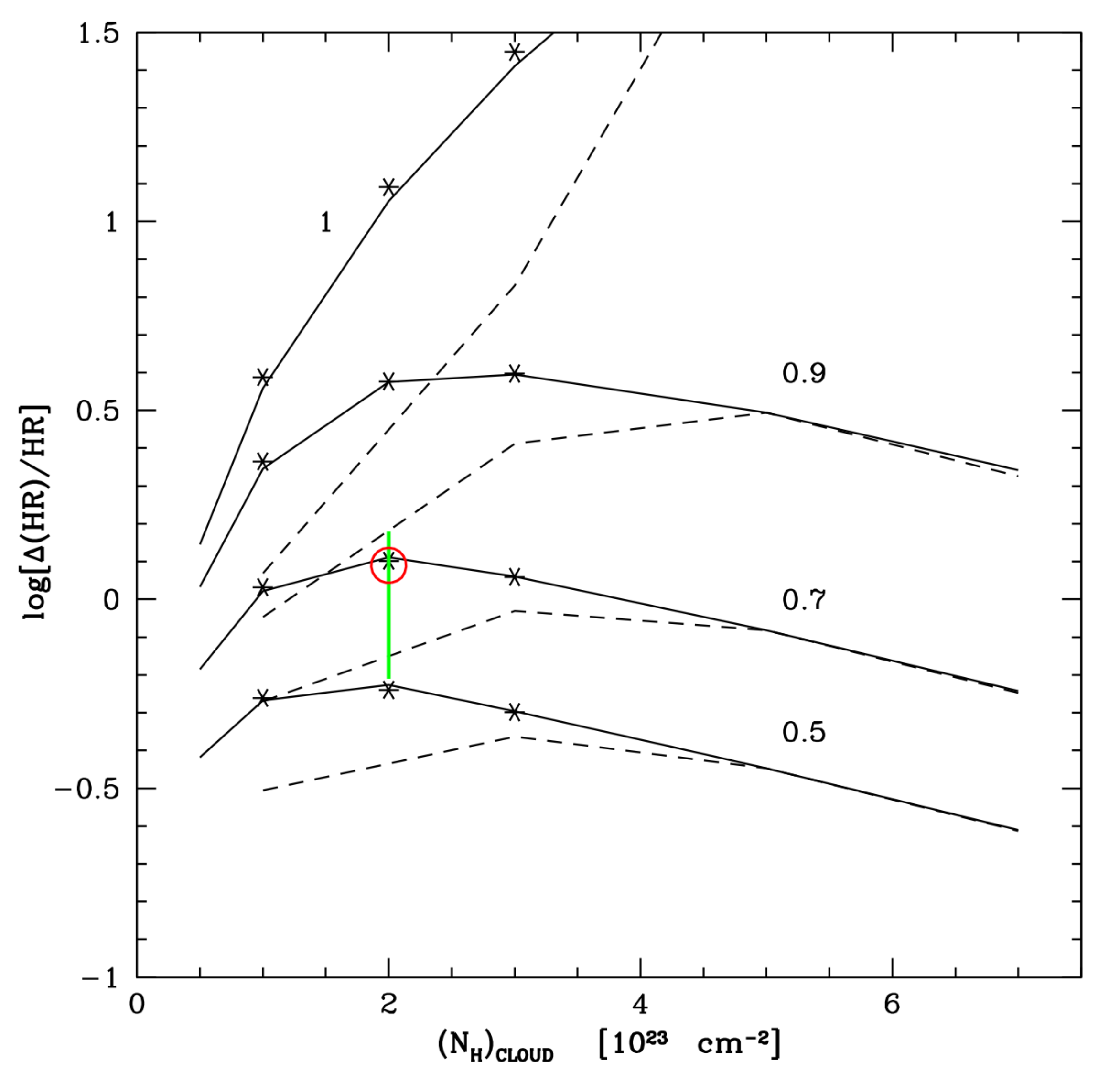} 
   \caption{ $\Delta(HR)/HR$  values obtained from synthetic {\em XMM} spectra as a function of occulting cloud column density, $N_H$,  for different
   values of  the covering factor and fixed $\gamma=2$ . Continuous lines refer to the case of unabsorbed Type~1 sources (see Eq.(3)); dashed lines refer to the case of Type~2
   sources  with uniform intrinsic absorption  $
 \tilde {N}  = 2 \times10^{23}$ cm$^{-2}$ (see Eq.(4)).
   Curves are labelled with the covering factor value. Asterisks show the same results for Type~1 sources with a different spectral index, namely 
   $\gamma=$2.3.
   The green segment indicates the range of values deduced from the spectral analysis for a specific observation of NGC~1365 (see text for details)
    and the red circle the value of  $\Delta(HR)/HR$ derived in this paper from the corresponding   $HR$ light curve.}
\label{HR_NH}
\end{figure}

As in the previous case the quantity which is more significant for the comparison with the observed light curves is the relative $HR$ variation  with respect to the unperturbed case, i.e. the case with $N_H=0$ for Type~1 sources or the case in which matter with  intrinsic column density $ \tilde {N}=2\times10^{23}$~cm$^{-2}$ (chosen as representative for the case
of  Type~2 sources) covers the whole source.  We have chosen a representative ``mean" value, $\gamma=2$, for the spectral index.
In these two cases,  $HR$ variations read
\begin{equation}
$$\big[{\Delta (HR) \over HR}\big]_{N_H}= {HR(2, N_H, C_F,0)-HR(2, 0, 1,0) \over HR(2, 0, 1,0)}$$
\end{equation}
and
\begin{equation}
$$\big[{\Delta (HR) \over HR}\big]_{N_H+  \tilde {N}}= {HR(2, N_H+  \tilde {N}, C_F,0)-HR(2,  \tilde {N}, 1,0) \over HR(2,  \tilde {N}, 1,0)}$$
\end{equation}
for Type~1 and Type~2 sources, 
respectively. Note that here and in the following  the covering factor $C_F$ always refers to
eclipsing cloud column density and not to the intrinsic column density in Type~2 sources.

In order to properly  ``simulate'' actual occultations, we must consider the possibility that only part of the X-ray source could be eclipsed by the obscuring cloud. This is expected if the eclipsing cloud is smaller than, or the same size as, the X-ray source, as it has  been already observed in the individual sources cited in the Introduction. 
Therefore, we repeated  our procedure for deriving synthetic quantities for several different values of the covering factor ($C_F$) to the central X-ray source. 

Figure \ref{HR_NH} shows  the synthetic quantity $\Delta(HR)/HR$  for Type~1 
and Type~2 sources
as a function of the eclipsing cloud column density, $N_H$,  for fixed $\gamma=2$,
 for the case of {\em XMM} instrument.  Continuous lines refer to Type~1 sources and show
how the synthetic quantity $\Delta(HR)/HR$ changes
in different  cases corresponding to  different covering factors of  the X-ray emitting source, ranging  from  the case of a complete eclipse ($C_F = 1$) down to a case in which half of the source is obscured ($C_F = 0.5$). For the case of Type~2 sources (shown as dashed lines) we take into account  an intrinsic constant absorption with  $  \tilde {N}= 2\times10^{23}$ cm$^{-2}$  plus the partial ($C_F < 1$)  to total ($C_F=1$)  cloud obscuration (see Eq.(4)). 

  Some tests, shown as asterisks in the figure,
 have been also done for the case of Type~1 sources with a different spectral index value, namely  $\gamma =$2.3. It is evident that 
 the change in the spectral index does not modify the results. The same analysis has been performed also for {\em {\em Suzaku}} and the results confirm those shown for {\em XMM}.

These results clearly show that $HR$ light curves with  significant and high humps ($\Delta (HR)/HR \ga  0.3$)  can be easily explained by clouds
crossing  the line of sight and eclipsing  at least half of the X-ray source ($C_F \ga 0.5 $).

 \subsection{Comparison with detailed spectral analysis}

The final result of the previous analysis, as it is apparent from the comparison of  Figures ~\ref{Dgamma}, \ref{refl} and \ref{HR_NH},
is that  large variations
in the $HR$ light curves cannot be due to either  spectral index variations or large flux variations, while interposing clouds with
different covering
factor can imply large $\Delta (HR)/HR $.

In order to better illustrate our analysis, and to test our conclusion in real cases, we analyze in more detail two observations for which we already performed a complete time-resolved spectral analysis. We choose the long 2005 {\em XMM-Newton} observation of Mrk~766 (Risaliti et al.~2011) where we found three clear occultations, due to three different eclipsing clouds, and the {\em Suzaku} observation of NGC~1365 reported in Maiolino et al.~2010. To compare the spectral analysis results with those  obtained with  our method, we must  take into account that in our analysis the hardness ratio variations are described by a gaussian curve and, hence,  each value we derive to quantify $HR$ humps
is related to this specific type of description. In practice, we can compare  the   $\Delta (HR)/HR $ values  derived here with those deduced from the spectral analysis  only on
time scales  corresponding  to 
intervals as long as   the gaussian half maximum  width  and  centered on  the maximum of the observed $HR$ hump. 

{\bf Mrk~766:} The observations during which the occultations occurred are the first and second in 2005 (Risaliti et al.~2011 Table~2). Our best fits include two highly significant events in the first observation, and one in the second observation.
 In these cases the statistics is good enough to allow  a detailed   analysis (Risaliti et al.~2011) of the $C_F$ and N$_H$ changes during each occultation, with a time resolution of $\sim$10~ks, therefore for each event we obtained a range of values for $C_F$ and N$_H$. We also ruled out a significant contribution from continuum variability: a fit allowing only for spectral index variability is ruled out at high statistical significance in each of the three events. Moreover, each event has been fitted with a model where all the components discussed in the previous Subsections (photon index, absorption, relative reflection/intrinsic flux ratio) were free to vary, and the best fit solution is compatible with constant photon index and variable absorption. In order to compare Risaliti et al.~2011 results with those of our method, as explained above,  we  must select in their Table 3 the time intervals corresponding to our gaussian curves,  i.e. 
$\Delta t(1)=[ 0-3\times10^4] $~s  and  $\Delta t(2)=[ 7\times10^4-1\times10^5] $~s for the first observation and $\Delta t(3)=[ 5\times10^3-2 \times10^4] $~s for the second observation.
The best fit values for covering factor and column density, obtained from spectral analysis for the three events, are N$_H$(1)=0.9-2.5$\times$10$^{23}$~cm$^{-2}$, $C_F(1)=0.82-0.89$; N$_H$(2)=1.7-1.84$\times$10$^{23}$~cm$^{-2}$, $C_F(2)=0.84-0.97$; N$_H$(3)=0.6-1.9$\times$10$^{23}$~cm$^{-2}$, $C_F(3)=0.60-0.85$.

If we compare these results with the  synthetic quantities  in Fig.~\ref{HR_NH},  
 for the  $N_H$ and $C_F $ intervals quoted above, we  expect:  [$\Delta(HR)/HR$](1)=1.5-4; [$\Delta(HR)/HR$](2)=2.2-10;  [$\Delta(HR)/HR$](3)=0.5-2.5. These  values must be compared  with  those derived from  the totally independent method of this paper, i.e. with the results of the gaussian fit of the hump in $HR$ light curves.  From Table~2 we find that the corresponding values are  [$\Delta(HR)/HR$](1)=2.15; [$\Delta(HR)/HR$](2)=2.36; [$\Delta(HR)/HR$](3)=1.03 in agreement with the above intervals.

{\bf NGC~1365}. The comparison described above for Mrk~766 has been repeated for the case of an obscured AGN,   NGC~1365 ($N_H\sim$2$\times$10$^{23}$~cm$^{-2}$). This specific case is
also shown in Fig.~\ref{HR_NH} as an example of the method we followed. The light curve for this same observation    is  the one shown in Fig.~\ref {examples}\,b 
 and shows a strong variability of $HR$ with a large peak around  $2\times10^5 $ s. A complete spectral analysis (Maiolino et al.~2010) demonstrated that in  the time interval  $[ 1.5\times10^5 -2.2\times10^5] $~s this variation is due to a cloud crossing the line of sight, with a column density of  $\sim 2 \times$10$^{23}$~cm$^{-2}$, and a covering factor $C_F =0.65-0.9$. These values are shown in Fig.~\ref{HR_NH} as a green segment connecting the intersection of the dashed curves with constant covering factor ($C_F =0.65$ and $C_F =0.9$)  for the fixed value  $N_H= 2 \times$10$^{23}$~cm$^{-2}$.
The corresponding range of $\Delta(HR)/HR$ is, from Fig.~\ref{HR_NH}, $\Delta(HR)/HR$=0.6-1.5, in agreement with our result of  Table~2 ($\Delta(HR)/HR$=1.23) shown as a red circle in the figure.

The   comparisons  described above clearly show that the analysis presented here provides results in agreement with complete spectral analyses. It is also apparent that  from our analysis it is not possible to  proceed in the opposite direction, i.e. derive two parameters ($N_H$ and $C_F $) with only one observable ($\Delta(HR)/HR)$.

\section{Results}
In the previous Section we showed that most of the variability of the X-ray hardness ratio, $HR$, observed in local bright AGNs can be due only to occultations of clouds crossing the line of sight.

Based on this conclusion, we can analyze the results in Fig.~\ref{diagnostic}, and qualitatively divide the variability events in four groups: \\
1) Low probability of chance fluctuation ($F_n \leq$10$^{-3}$), and $\Delta(HR)/HR> 0.1$. These are events whose interpretation is unambiguous, and we expect to be able to confirm the occultations in all cases through time-resolved spectral analysis.\\
2) Low  $F_n$,  ($F_n\leq$10$^{-3}$), and $\Delta(HR)/HR<0.1$. These are cases  in which the variability is not due to fluctuations, but the interpretation is less certain. A strong change of photon index, or an extreme flux variation, may produce the same effect as an occultation.\\
3) $F_n>$10$^{-3}$, and $\Delta(HR)/HR> 0.1$. These are candidate occultations, for which the statistical significance of the $HR$ variation is not high enough to rule out a statistical fluctuation.\\
4)  $F_n>$10$^{-3}$, and $\Delta(HR)/HR<0.1$. These are marginal events, which cannot be confirmed through a spectral analysis due to low statistics, and where the spectral variation is modest. We expect that a fraction of these events are occultations (probably with a low covering factor), but we will conservatively assume that no absorption variation has been detected in these cases.  

We note that the limiting values of $\Delta(HR)/HR$ and $F_n$ are arbitrary, and rather conservative. Indeed, events with $F_n$ of a few 10$^{-3}$ are still quite significant (the customary 3-$\sigma$ limit corresponds to $F_n=3 \times$10$^{-3}$), and our analysis of the photon index variability suggests that most of the events with $\Delta(HR)/HR> 0.1$ are due to occultations.
However, we adopt this conservative approach based on practical considerations: 1) the value $F_n =10 ^{-3}$ roughly corresponds to the limit where, based on our past experience,  we are sure to be able to confirm the results through time-resolved spectroscopy; 2) the value $\Delta(HR)/HR> 0.1$ is a safe upper limit;  at any rate  we plan to perform time-resolved spectroscopic analysis of the cases with low $F_n$ and $\Delta(HR)/HR<0.1$, in order to complement the present analysis based only on light curves   and to test if the thresholds   adopted here  for $F_n$ and $\Delta(HR)/HR$ are confirmed.

In summary we have 15 out of 42 sources with highly probable occultation events ($\Delta (HR)/HR > 0.1$ and $F_n\leq 10^{-3}$), 3 sources with $\Delta(HR)/HR<0.1$ and $F_n\leq 10^{-3}$, i.e. cases with highly significant variability events, which may be due either to occultations or to intrinsic variability, and  24 sources with no strong evidence for spectral variability. From these numbers we conclude that at least 1/3 of local AGNs shows  occultations on time scales of the order of, or shorter than, the observing time.

We can further refine this analysis by comparing our results with the physical occultation time scales of each source, under the hypothesis that the observed spectral variations are due to broad line clouds crossing the line of sight.

We want to give an estimate of the eclipse duration timescale for each of the 
sources listed in Table~1. The simplest and most effective (in terms of source coverage due to the intervening cloud) 
geometry for the eclipse event is the one in which, assuming for simplicity both source and cloud as spherical, the cloud centre 
moves on a trajectory that intercepts the source centre; in this case, following Risaliti et al. (2007), the eclipse timescale can be 
evaluated as  $2(R_X+R_{cloud})/v_{cloud}= 2R_X/v_{cloud}(1+ R_{cloud}/R_X)$, where $R_X$ is the X-ray source radius, $R_{cloud}$ is the radius of the cloud, and $v_{cloud}$ is the cloud velocity.
 However, 
in general we must expect  a broad range of different geometrical configurations due both  to varied cloud paths and  to different size of the 
clouds themselves.
In fact, in our framework, still maintaining the simplifying hypothesis of spherical geometry for source and clouds, we expect clouds to cross the line of sight to the source, and obscure it, moving on different trajectories (that we suppose locally rectilinear for simplicity), thus allowing for a distance, $d$, between the trajectory of the moving cloud centre projected on the plane of the sky and the source centre, 
$d= f_dR_X$ , where $f_d$ is a non-dimensional parameter, whose values must be $0\le f_d<(1+ R_{cloud}/R_X)$ in order to have an eclipse event at all. 
This of course affects 
both the resulting duration of the eclipse and the corresponding maximum covering factor, $(C_F)_{max}$, attained during the passage.
As for the size of the eclipsing cloud, that we parameterize with its radius $R_{cloud}$, any assumption on that just aims at defining an order of magnitude evaluation; at the present stage, only upper  and lower {\it operative} limits exist for the cloud extension.
In fact, eclipsing  clouds are detectable with our method only if  their physical and geometrical properties are such to
produce an observable emission decrease and hence a hump in the $HR$ light curve.  Too
large clouds would  not be detectable, since their passage would produce too long eclipsing times which could result longer than the available light curve  duration (see below).
On the other hand,  too small clouds  would leave  the source emission almost unaffected.  Ultimately, based on our analysis of model convolution with response matrices of the instruments, we  find that clouds  with detectable effects on the source emission must produce  maximum covering 
factors larger than 0.1  and this implies  it must be  $R_{cloud}\geqslant 0.3 R_X$. 
Taking into account the considerations above, the timescale for the eclipse duration,$\Delta t_{ecl}$, that we want to evaluate as the time a BLR cloud takes to cross the X-ray source (Risaliti et al., 2007), can be expressed as

$$\Delta t_{ecl} \simeq 2~ Q~ {R_X\over v_{cloud}} , $$
where $Q \equiv Q\left( {d\over R_X}, {R_{cloud}\over R_X}\right)$ is a factor that depends on the values of $d/R_X$ and $R_{cloud}/R_X$.
In order to define the most representative estimate of the factor $Q$, we have explored the portion of parameters space identified by 
the ranges $0.3\le R_{cloud}/R_X \leq 1$ and $0\leq d/R_X\leq (1+R_{cloud}/R_X)$, with the constraint that the eclipsing event produces a maximum covering factor $\ge 0.1$, {\it i.e.}, an observable variation effect in the hardness ratio light curve. From  this analysis, we conclude that 
 $Q\simeq 1.5$ can be chosen as an average,   representative value  for the resulting range, and, as a consequence, we adopt this value for our
 estimate of the duration of the source obscuration in the cloud passage, that is the eclipse timescale 
 $$\Delta t_{ecl} \simeq 3~{R_X\over v_{cloud}}\simeq  7.5 R_S/ v_{cloud} ,$$
  where $R_S= 2GM_{BH}/c^2$ is the Schwarzschild radius and we have assumed the source diameter
 $D_X=2R_X\simeq 5 R_S$. This last assumption is supported also by the recent results of a detailed study 
 of one of the sources in our sample, namely IGR~J21277+5656 (also called SWIFT~J2127.4+5654),
 by Sanfrutos et al. (2013); in this work,  time resolved spectral analysis of the source, interpreted as a partial eclipse by an intervening 
BLR  cloud-like absorber, enables the authors to constrain the size of the X-ray emitting region, that turns out to be   
$D_X\leq 10.5~ GM_{BH}/c^2 =  5.25 R_S$, when adopting $M_{BH} \sim 1.5\times10^7 M_{\sun}$, a value estimated from single-epoch optical spectroscopy.

\medskip
For cloud velocities two different  estimates have been compared.

{\it  i)} In Type~1 sources, cloud velocity  can be directly  estimated from the broad line widths.  Among  the  average speed and dispersion values found in quasars of the Sloan Digital Sky Survey (Ahn et al.~2012) we note that more than 90\% of the SDSS quasars have line widths within a 2000-6000~km/s range, i.e. with a smaller dispersion with respect to the central value than the uncertainty on the black hole mass and/or on the size of the X-ray source.  Therefore,
we have assumed v$_{cloud}$=4000~km/s, 
as  representative cloud speed for all  sources, except for the three  NLS1 sources for which $v_{cloud} =1000$~ km/s has been assumed.

{\it  ii)} The cloud velocity can be estimated  also in
 the hypothesis that X-ray obscuring clouds undergo a keplerian motion around the central black hole, i.e. $v=(GM_{BH}/R_{BLR})^{0.5} $, 
  and that the Bentz  et al. (2013)  radius-luminosity correlation, $R_{BLR} \simeq 33.6~ [L_{5100}/10^{44}]^{0.533} $ light~days,    holds for
  every type of source in our sample. The use of the above expression needs an estimate of the   $5100 \AA$ luminosity, $L_{5100}$, which, for several sources in our sample, is not available from observations and must be in some way  evaluated.  Our choice was to adopt the values of $L_{5100}$ present in literature, when possible (Bentz et al. 2013, Winter et al. 2010), and, otherwise,  derive  $L_{5100}$ by means of the optical  to X-ray  spectral index derived by Young et al.~(2010),  where the monochromatic  2~keV flux has ben obtained by 
   extrapolating  the measured 15-195 keV flux from the Swift/BAT 54 month catalogue (Cusumano et al.~ 2010) using the spectral index given in the same catalogue.
  From the   value   of $L_{5100}$ and using the  relationships  quoted above, 
   an approximate value for the cloud
 speed and hence for the eclipsing time can be  derived for each of the sources in Table 1, except  for  2MASX J04532576+0403416, for which no value
 for the central black hole mass is available in the literature.

We find that  the eclipsing time values  evaluated by  using $L_{5100}$, taken from literature or estimated as described  above,  
and the values obtained assuming a constant cloud velocity  are of the same order of magnitude, since the ratio of the two estimates  is  in the range 
[0.3-4]   for
all the sources in our sample (apart from a couple of  exceptions, MCG~06-30-15 and ESO~141+55, for which the ratio is $\sim 5$) and for  about 80\% of the sources it is in the narrower  range [0.4-2].
 In the following we always adopt 
for the estimate of the crossing time $\Delta t_{ecl}$  the luminosity-based method and the derived values are listed in Table~1 for
each source of our sample.

\medskip
 
\begin{figure}
\includegraphics[width=8.5cm]{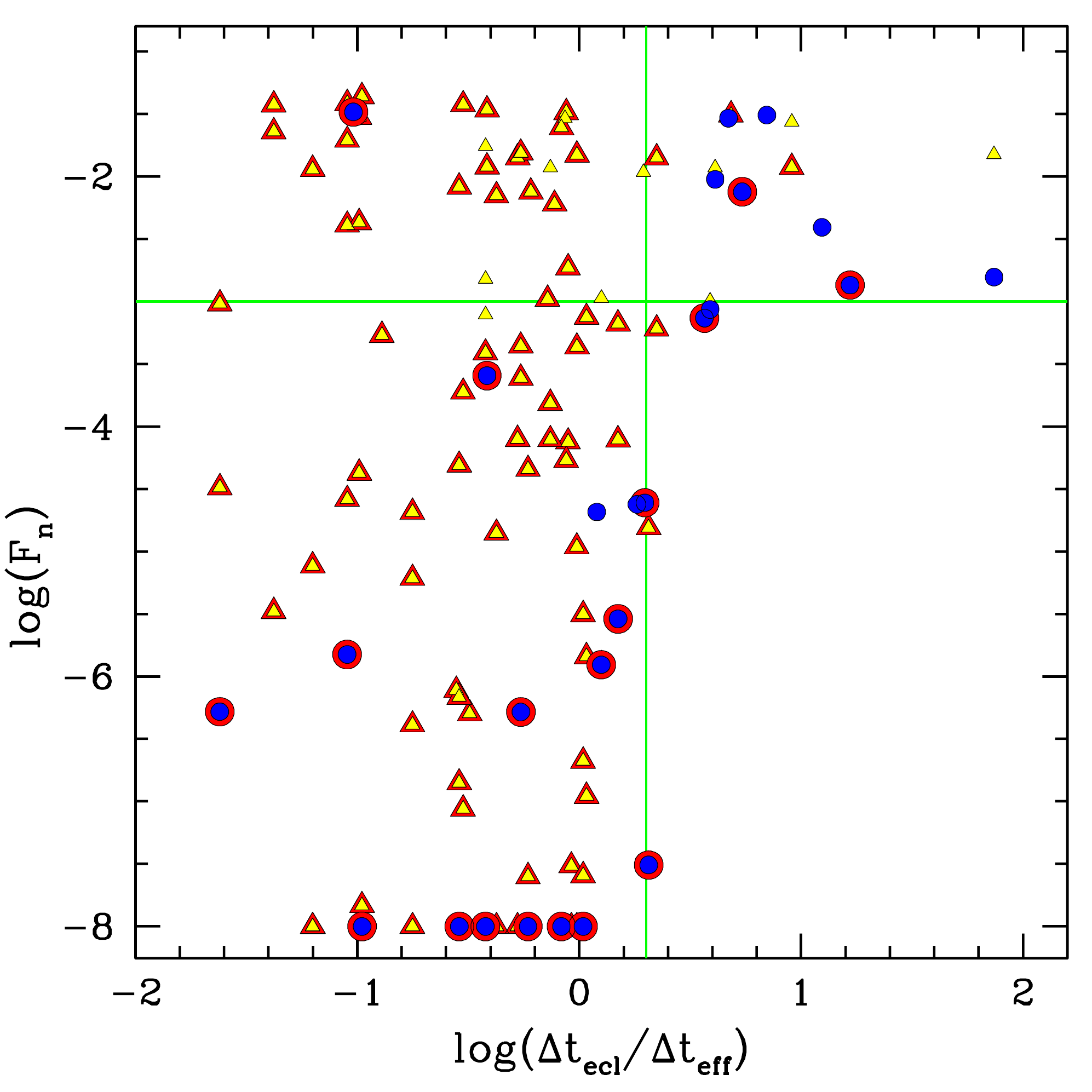} 
\caption{Statistical significance of the hardness ratio variations, $F_n$ as a function of $\Delta t_{ecl} / \Delta t_{eff}$, where $\Delta t_{eff}$ is the observation length, and  $\Delta t_{ecl}$ is the duration of the eclipse estimated assuming that the $HR$ variation is due to a broad line cloud covering the source  (see text for detail). Each event is represented by a triangle. Circles indicate the most significant event for each source in the sample. Big, red contours are for events with $\Delta(HR)/HR>0.1$. Note that for peaks with  significance $F_n< 10^{-8}$ the value $F_n= 10^{-8}$ is shown in the figure. 
  }
   \label{Pn_tempi}
\end{figure}

 The estimated $\Delta t_{ecl}$ 
can be correlated 
with the results of our light curve analysis. In Figure~\ref{Pn_tempi} we plot the probability $F_n$ for each event, versus the ratio, $\Delta t_{ecl}/ \Delta t_{eff}$,  between $\Delta t_{ecl}$ and  the effective duration of the observation, $\Delta t_{eff}$. 
 We must stress here that, once we have performed our fit on a given $HR$ light curve, we do have an evaluation of the timescale for each variation we fitted with a gaussian shape, defined by the parameter measuring the gaussian width. In doing the fit, the effective duration of each single observation poses a limit on the width of the fitting functions we use to match the observed  hardness ratio variations in the corresponding light curve: we can reliably fit only variations that are clearly visible within the duration of the light curve, {\it i.e.}, those occurring on a timescale that is $\leq  2\Delta t_{eff}$. For larger values, we could in principle still be able to measure significant spectral variations, however our method requires at least $\sim$half of an eclipse to be monitored in order to have a significant gaussian component in our fits of the light curves. In the framework of our physical interpretation, our estimate of the eclipse timescale $\Delta t_{ecl}$ for a given source should be representative of expected timescales for $HR$ variations, although we may envision significant deviations of the actual observed variation timescales, due to the simplifying assumptions in our calculation above. Keeping in mind these considerations, the plot in Fig.~7
shows the two most relevant results of our work.

 1) We assume that the operative limit on observable and analysable hardness ratio variations timescale ($\leq  2\Delta t_{eff}$)  can be extended also to our estimate of a typical eclipsing time for a source, $\Delta t_{ecl}$, and we trace a vertical line for $\Delta t_{ecl}/ \Delta t_{eff}=2$ in Fig.~7.
 This figure shows than no highly significant ($F_n\leq 10^{-3}$) events are detected with $\Delta t_{ecl}/ \Delta t_{eff}>2$. We point out that, despite the large uncertainties in our estimates, the quantities involved in this comparison, i.e. the statistical significance of the variations, the estimated BLR eclipsing times, and the actual observing times, are both observationally and physically unrelated. Therefore, a clear result such as the lack of events in the bottom-right part of the plot in Fig.~\ref{Pn_tempi} represents a   strong self-consistency check both of  our eclipse-based interpretation of the spectral variations, and of  our hypothesis that such eclipses are due to BLR clouds. 
\\
In order to clarify the relevance of the result shown in Fig.~7, it is useful to mention that we analysed the distribution of values of the ratio $\Delta t_{ecl}/ \Delta t_{eff}$ estimated for all the observations in our sample and we found this distribution to be almost symmetric with respect to $\Delta t_{ecl}/ \Delta t_{eff}\simeq 1$; more specifically,  we found that the number of observations corresponding to $\Delta t_{ecl}/ \Delta t_{eff}\geq 2$ is 
about 39\% of the total number of observations;
as a consequence, the absence of significant $HR$ variation events in the lower right  portion of the plot in Fig.~7 cannot be explained with a scarcity  of observations with $\Delta t_{ecl}/ \Delta t_{eff}\geq 2$.
In fact, the only events  we have been able to fit for observations with $\Delta t_{ecl}/ \Delta t_{eff}> 2$ and  appearing in the plot are located in the upper right portion, since, according to our method of analysis, their significance is low.
It is important to point out that the absence of statistically significant $HR$ variation events in observations with $\Delta t_{ecl}/ \Delta t_{eff}\geq 2$ is not a truism, since our procedure would in principle allow for the detection of significant events in a $HR$ light curve, even though for that same observation $\Delta t_{ecl}\geq 2 \Delta t_{eff}$. In fact, the only limit to detect an event would be on the actual duration of the event itself, that is on the width measure of the appropriately  fitting gaussian component:  we 
can in principle  always fit  $HR$ variation events 
whose duration measure ($\simeq w_n$, see Section~3.3) 
 is $\leq 2 \Delta t_{eff}$, 
 but,  for an observed 
light curve with $\Delta t_{ecl}\geq 2 \Delta t_{eff}$,
 in those cases it would be 
$w_n <  \Delta t_{ecl}$ or even $\ll \Delta t_{ecl}$.
No such statistically significant events turn out to exist from our analysis and this is right what one would expect if (and only if) the origin of significant $HR$ variations were indeed due to a cloud-like structure eclipsing the source while crossing our line of sight to the source itself.

2) From an analysis of the left part of the plot (i.e. the only region of the parameter space where it is possible to see occultations due to BLR clouds), $\Delta t_{ecl}/ \Delta t_{eff} <2$,  we find 16 sources with at least one highly significant event ($F_n\leq $10$^{-3}$). Out of this subgroup, only two objects have $\Delta(HR)/HR<0.1$, thus not excluding the   possibility of   intrinsic spectral variability. The upper part of the diagram ($F_n>$10$^{-3}$) contains the most significant event for only one source; in addition   other 6 sources exist in our sample having  $\Delta t_{ecl}/ \Delta t_{eff}<2$ (not shown in the figure since they  do not have any significant hump characterized by $F_n < 0.05$
in their $HR$ light curves).  Therefore the fraction of sources for which is  $\Delta t_{ecl}/ \Delta t_{eff}<2$ and  having detected X-ray occultations is 14/26=54\%. This confirms that eclipses by BLR clouds are common among AGNs.\\
 This estimate should be regarded as a lower limit, for several reasons.\\
{\it a)} Our analysis method is relatively fast and homogeneous for the whole sample. However, a more {\it ad-hoc} analysis of individual light curves could reveal more significant variability events. Moreover, the significance of the possible column density variation, as measured in a time-resolved spectral analysis, is of course related to the significance of the hardness ratio variation in the light curve, but with some dispersion. This is due to the degeneracy among model parameters in the spectral fits, which are dependent on the details of the model and on the data quality, and should be estimated on a case-by-case basis. Our experience suggests that the value $F_n$=10$^{-3}$ is a reasonable threshold: events with a higher statistical significance typically are confirmed by a complete spectral analysis, while in less significant ones the variations of model parameters such as column density and continuum spectral index are degenerate. \\
 {\it b)} Significant variability events may be detected comparing the spectral analysis of {\it different} observations.
A known example is the Type~2 Seyfert Galaxy NGC 7582, for which  the hypothesis of obscuring clouds has been supported by  the comparison  of 
two  XMM-Newton observations  taken four years apart (Piconcelli et al. 2007)  and  by  comparing the spectral analysis of a series of short {\em Suzaku}  observations (Bianchi et al.~2009).  
On the contrary, for this source the only observation meeting the criteria of our sample  and analyzed with our method does not show any significant variability.\\
{\it c)} Occultations by BLR clouds may occur, but we could have missed them in our observation sample. This is for example the case of NGC~4388, a Type~2 Seyfert galaxy for which   all the  existing {\em XMM} observations do not enter in our sample since last less than 80 ks (our lower limit), and  the analyzed {\em Suzaku} observation  does not show any variability since it is well fitted by a constant 
($\chi^2_0=0.99$)
Hence,  for this source we  do not detect 
any significant variability whereas variability  has been observed in the data obtained by a different instrument  when the source was in a Type~1 unobscured state during a Rossi-XTE monitoring campaign (Elvis et al.~2004). It is interesting to notice that some of our most significant occultation events have been detected in the NLS1 Mrk~766 thanks to a particularly long {\em XMM-Newton} monitoring (6 orbits), but with a small fraction of the observing time affected by these events. Therefore, it is likely that a single $\sim$100~ks long observation (as in most of our sources) would not detect any occultation in this source, even if it is one of our best cases.\\

\section{Conclusions}

The aim of this paper was to test on a larger sample the hypothesis of X-ray occulting clouds already proved for a small number of sources [see Risaliti~(2010) and references therein] in recent years.
Therefore, we have selected  a statistically representative sample  of bright X-ray AGNs for which long lasting 
 {\em XMM}  ($>80$ ks) and {\em Suzaku} ($>50$ ks) observations are available
and, for each suitable observation, we have derived the hardness ratio ($HR$) light curve.  In these $HR$ curves we have  identified  any time interval where the $HR$  is not
constant but shows a hump and we have  fitted these humps with gaussian curves in order to have a way to quantify
the hump location, width and height  over the constant $HR$ level. We have also computed the statistical significance of the obtained  fits, deriving for each gaussian curve its null probability, $F_n$ from the F-test.
 The results of this analysis are shown in Table 2 and constitute the basis for the   further analysis.

It is well known that the possible interpretations of humps in $HR$ light curves as due to occultations by interposing clouds must be confirmed through a complete spectral analysis, as  has  been done in the above reported cases. However, 
the idea of this paper was to devise a method which could allow a first concise but faster analysis on a large number of sources. For this reason we have investigated the dependence of $HR$ humps (in particular the height of each hump over the constant level represented by the fractional  variation, 
$\Delta (HR)/HR$ ) on the cloud column density and covering factor and on the X-ray radiation spectral index.  To quantify these dependences we have employed  synthetic models deriving (see Figs.~\ref {Dgamma}  to \ref{HR_NH})  a specific limit :  humps with $\Delta (HR)/HR >  0.1$  cannot be due to spectral index variations but can be explained with the effects of crossing clouds. With this limit we can already say that a large number of the detected humps is originated by crossing clouds.
  This conclusion can be strengthened by the analysis of the theoretically deduced eclipsing times. In fact, for each source in the sample, we have  computed the time necessary  for  a cloud with a diameter  the order of $5~R_S$  and located at BLR distance to cross the central  source  and  we have compared this time  to each observation length.

  The main results of our analysis, shown in Fig. ~\ref{Pn_tempi}, are the following.   
  First, it is evident  that there are not  significative humps ($F_n \leq 0.001$) in the region corresponding to  theoretical
 eclipsing times longer than  twice the observation time.  
  Since the choice of the physically relevant time-scale ($\Delta t_{ecl}$) appearing in the time ratio
 used for Fig.~\ref{Pn_tempi} follows from our eclipsing cloud interpretation of $HR$ variability,
 the empty  lower right region in Fig.~\ref{Pn_tempi}
 is a strong confirmation of this same  physical interpretation. In fact, 
 no other physical source of $HR$ variability would imply a distribution of significant $HR$ time variation events showing a dependence on the timescale for a BLR cloud to cross the X-ray source extension. 
 
 Secondly,  Fig.  \ref{Pn_tempi} allows an evaluation  of  X-ray occultation frequency. Eclipses  by BLR clouds seem very common in AGNs since in our sample the fraction of sources with   good candidate   X-ray occultations  detected is 54\%,  among the sources for which $\Delta t_{ecl} \la \Delta t_{eff}$ .
 
In conclusion, our study of a statistically representative sample of AGNs prompts the validation of a method for the identification of BLR cloud occultation events based on the analysis of 
the $HR$ light curves, 
and the results we obtain give full support to the  physical interpretation of $HR$ time variability
in terms of X-ray source occultations by BLR-like clouds crossing the line of sight to the source itself.

\section*{Acknowledgements}
This work has been partly
supported by grants  NASA NNX08AN48G,  PRIN-MIUR 2010-2011  "The dark Universe and the 
cosmic evolution of baryons: from current surveys to Euclid "  and
 PRIN-INAF 2011 "Black hole growth and AGN feedback through the cosmic time".

\end{document}